\documentclass[aps,pra,twocolumn,groupedaddress,superscriptaddress,bibnotes,amsfonts,longbibliography,citeautoscript,a4paper,10pt]{revtex4-2}
\usepackage{latexsym}
\usepackage{amsmath}
\usepackage{amssymb}
\usepackage{graphicx}
\usepackage[T1]{fontenc}
\usepackage[open]{bookmark}
\usepackage{hyperref}
\usepackage{orcidlink}
\hypersetup{colorlinks=true,allcolors=blue}
\usepackage{xcolor} 
\usepackage{csquotes}
\usepackage{bm}

\setcitestyle{numbers,super,comma}

\definecolor{refblue}{HTML}{0070C0}

\providecommand{\pnl}[1]{{\textcolor{black}{(#1)}}}

\begin{document}

\title{
Time- and frequency-domain study for electron beams penetrating dielectric nanospheres: fingerprints of Cherenkov and transition radiation}

\author{Wenhua~Zhao\,\orcidlink{0009-0004-5721-607X}}
\email{Corresponding author: wzhao@physik.hu-berlin.de}
\affiliation{Max-Born-Institut, 12489 Berlin, Germany}
\affiliation{Humboldt-Universit\"at zu Berlin, Institut f\"ur Physik, AG Theoretische Optik and Photonik, 12489 Berlin, Germany}
\author{Christos~Tserkezis\,\orcidlink{0000-0002-2075-9036}}
\affiliation{POLIMA---Center for Polariton-driven Light-Matter Interactions, University of Southern Denmark, 5230 Odense M, Denmark}
\affiliation{Danish Institute for Advanced Study, University of Southern Denmark, 5230 Odense M, Denmark}
\author{N.~Asger~Mortensen\,\orcidlink{0000-0001-7936-6264}}
\affiliation{POLIMA---Center for Polariton-driven Light-Matter Interactions, University of Southern Denmark, 5230 Odense M, Denmark}
\affiliation{Danish Institute for Advanced Study, University of Southern Denmark, 5230 Odense M, Denmark}
\author{Kurt~Busch\,\orcidlink{0000-0003-0076-8522}}
\email{Corresponding author: kurt.busch@physik.hu-berlin.de}
\affiliation{Humboldt-Universit\"at zu Berlin, Institut f\"ur Physik, AG Theoretische Optik and Photonik, 12489 Berlin, Germany}
\affiliation{Max-Born-Institut, 12489 Berlin, Germany}

\date{\today}

\begin{abstract}
We present a theoretical study of Cherenkov and transition radiation for swift electron beams penetrating dielectric nanospheres using material models of different sophistication. Specifically, we perform a combined time-domain (numerically, via the discontinuous Galerkin time-domain method) and frequency-domain (numerically and analytically, via Mie-based theory) study, including the induced-field distribution, cathodoluminescence (CL) multipole/directional decomposition, as well as the time-dependent angular power flow. For low velocities below the Cherenkov threshold, we show that transition radiation is dominant in the far-field CL, and the near-fields at the transition points are primarily responsible for the main features observed in the far-field. For higher velocities far beyond the Cherenkov threshold, we identify the fingerprints of the observable Cherenkov front. Specifically, a constant-permittivity model allows us to isolate the respective contributions of  CR and TR to the far-field radiation, thereby facilitating the interpretation of the results for a more realistic material model that includes material resonances. Our combined time- and frequency-domain framework provides a direct view of radiative excitation channels for swift electron beams penetrating dielectric nanoparticles, thereby revealing their interplay beyond the conventional frequency-domain analyses.\\
\noindent \textbf{Keywords:} Nanophotonics; cathodoluminescence; Mie theory; Poynting flux; Cherenkov radiation; transition radiation
\end{abstract}

\maketitle

\section{Introduction}
\label{sec:introduction}

The free electron beam provides a well-localized, broadband, and sub-wavelength excitation source, which couples to photonic modes via its evanescent near-fields and therefore plays a fundamental role in  nanophotonics~\cite{paper149,Polman_2019,talebi2017interaction,Abajo2021}. This includes electron energy-loss spectroscopy (EELS)~\cite{egerton2011electron,Ritchie1988,paper149}, which provides insights into both bright (such as lower-order Mie-type resonances) and dark photonic modes (such as longitudinal bulk modes), and cathodoluminescence (CL)~\cite{Polman_CL_2016,CLreview2023}, which captures the radiative part of the excitations. On the one hand, electron beams are widely used to study plasmonic excitations in metallic nanoparticles (NPs) containing free conduction electrons~\cite{paper369,hohenester2009electron,paper085,paper149}. In particular, the evanescent nature of the electron source allows efficient coupling to surface plasmon polaritons (SPPs) that lie outside the light cone in extended nanostructures, localized surface plasmons (LSP) in nanoparticles, and longitudinal bulk plasmons (BP), which are generally optically dark.
On the other hand, in dielectric nanostructures, the electron beam drives polarization currents and excites Mie-type resonances, enabling the investigation of multipole modes, transition radiation (TR)~\cite{TR_Ginzburg1996, TR_Chen2023} and Cherenkov radiation (CR) when the Cherenkov condition is fulfilled~\cite{Analytics_CR_TR_2003,TR_Ginzburg1996}.
More broadly, electron-induced radiation provides a powerful tool for nanophotonics. For example, Sapienza \emph{et al.}~\cite{Sapienza2012} demonstrated deep-subwavelength mapping of photonic modes using angle- and spectrally resolved CL.
The central question addressed in this study is how swift electron beams excite photonic modes in dielectric NPs, using silicon as an example, and how these excitations manifest themselves in the corresponding optical response.

Silicon (Si) is a widely studied material~\cite{SiliconSpringer} and one of the central materials in nanophotonics, which combines technological maturity with rich optical properties~\cite{Si_opticalProperties2025,Si_Tarrio1993}. 
Si is intrinsically low-loss in the near-infrared and Si NPs exhibit strong Mie-type excitations in the visible range due to their high refractive index and geometry, which render them efficient resonators and enable efficient nanoscale field confinement and tailored scattering. In the ultraviolet, Si even offers an interband plasmonic response~\cite{Dong2019}.
When probed by swift electrons, Si NPs further reveals a complex excitation spectra, accessible through EELS and CL. In addition to geometric Mie resonances, swift electrons can couple to interband transitions, bulk and surface excitations, and hybrid resonances that arise from the interplay between Mie resonances and the electronic band structure~\cite{TserkezisHybridisation}. Moreover, for penetrating trajectories, swift electrons can excite TR~\cite{TR_Ginzburg1996, Fiedler2021,Pogorzelski1973} and, when permitted by the refractive index condition, CR~\cite{TR_Ginzburg1996,Pogorzelski1973,paper149} inside the Si NPs. Therefore, EELS and CL of Si NPs provide a powerful, nanoscale resolved probe of both their optical resonances and material-specific electronic structure and, in this study, we focus on analyzing CL spectra of Si NPs.

At lower electron velocities which are below the Cherenkov threshold, TR dominates the power radiated into the far-field and can build characteristic interference patterns~\cite{Fiedler2021} in which the dipolar contribution~\cite{ebel_nanoph14} is most prominent. At higher electron velocities, beyond the Cherenkov threshold, the analytical treatment~\cite{Analytics_CR_TR_2003} shows that, for a dielectric sphere, CR and TR are inseparably mixed within multipole coefficients. The evaluated angular spectral radiation intensity differs from the traditional Tamm formulation~\cite{TR_Ginzburg1996} for planar interfaces as the surface-curvature induces a mixing of the radiation channels. 
In a broader context, García de Abajo \emph{et al.}~\cite{Abajo2004_Boundary_effects_CR} showed that dielectric boundaries modify CR through reflection and interference, while interface crossings generate TR-like emission and finite geometries couple the radiation to guided or resonant modes.

Pioneering TEM-CL experiments by Yamamoto and co-workers~\cite{Yamamoto01091991,Yamamoto1996_CR_TR_thin_plate,Yamamoto1996_TR_imaging} observed TR and CR from thin films and particles. 
Subsequent experiments~\cite{StoegerPollach} measured the TR excitation probability using both optical spectroscopy and EELS, providing a quantitative basis for assessing its contribution to EEL and CL spectra.
More recently, Scheucher \emph{et al.}~\cite{Haslinger2022} separated coherent and incoherent contributions of CL in time-resolved coincidence measurements.
Electron-beam studies have further enabled nanoscale mapping of excitation pathways~\cite{Varkentina2022}, chiral emission~\cite{Sannomiya2021},  whispering-gallery modes~\cite{Sannomiya2023,Sannomiya2025}, and directional emission via control of the Kerker effect.~\cite{soler_acsphot12} Recent measurements on silica microspheres have further disentangled CL generation mechanisms~\cite{Aharon2026}. 
Here, we complement these studies by resolving the time-dependent interplay of TR and CR for penetrating electron trajectories.

We perform a comprehensive study of the aforementioned effects in a Si spherical NP from both time-domain and spectral/angular perspectives. We examine its CL response using various material models, from a simple non-dispersive dielectric model with constant permittivity to a fully dispersive Drude--Lorentz material model~\cite{Si_lorentz2017,DrudeLorentz2024}.
To shed further light on this intricate interplay among different physical mechanisms, we perform numerical computations by solving Maxwell's equations using the discontinuous Galerkin time-domain (DGTD) finite-element method (FEM)~\cite{DGTD,Elli_Wenhua2023,Kiel_metalspitze}, which facilitates the spatio-temporal resolution of the excitations in the nanometer and femtosecond range (for details, see Sec.~\hyperref[sec:methods]{Methods}.
A recent study~\cite{Agwire2026} has shown that time-dependent analyses of surface plasmon polaritons in silver nanowires can provide valuable information and deeper understanding of interaction of electron beams with NPs. Combining time-domain simulations with analytic frequency-domain calculations is thus introduced as a comprehensive treatment for complex nanophotonic systems under the electron microscope.

\section{Theory} \label{sec:theory}

In this section, we first briefly summarize the underlying material models 
for the silicon sphere under study. We then revisit the theory of cathodoluminescence (CL), introducing the angle-resolved Poynting flux and CL emission probability. Finally, we recall the physical mechanisms of Cherenkov and transition radiation for swift electrons penetrating nanoparticles.


\subsection{Material models} \label{subsec:material_model}
To facilitate our understanding, we adopt two different material models for Si. The first one, which makes the underlying physics more transparent, is based on a constant permittivity $\varepsilon=16+0^+\mathrm{i}$ with negligible loss. This model does not incorporate material dispersion, and the Cherenkov behavior is much more straightforward to observe and analyze, so that all excitations can be unambiguously attributed to geometry 
or interference. 

\begin{figure}[h]
    \centering
    \includegraphics[trim = 80mm 85mm 75mm 45mm, clip, width=0.5\textwidth]{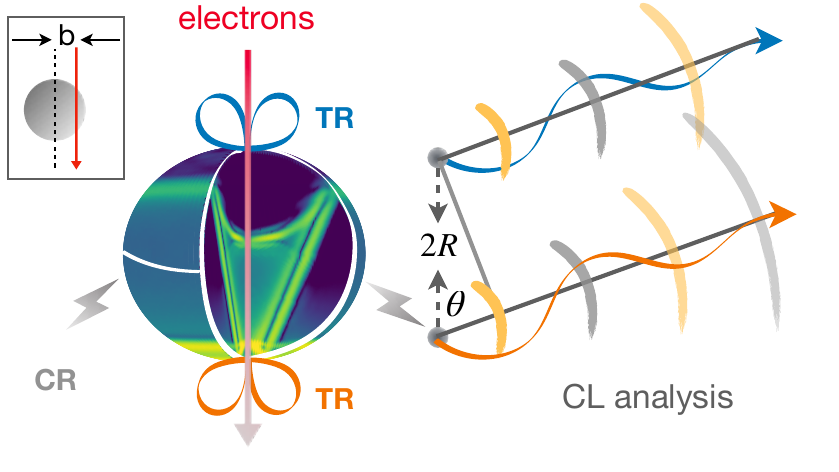}
    \caption{\textbf{Illustration of the interaction between a spherical nanoparticle (NP) 
    and a swift electron beam, highlighting the excitation of Cherenkov 
    radiation (CR) and transition radiation (TR).
    } 
    The electron beam penetrates the NP with radius $R$ at zero impact parameter $b$. Emitted light is collected in the far field, where radiation originating from the two crossing points interferes at angle $\theta$, which denotes the polar angle with respect to the $z$-axis. As the electron velocity exceeds the Cherenkov threshold, a CR cone forms inside the sphere and reaches the boundary, where it is partially transmitted and contributes to the cathodoluminescence (CL) signal. Simulation parameters: $\varepsilon = 16$, $\beta = v_e/c = 0.7$, $b = 0$. 
    }
\label{fig:Fig_1}
\end{figure}

In reality, like any causal material, Si exhibits both loss and dispersion, which strongly modify the supported excitations. Therefore, we eventually need to expand the study to a realistic description based on experimental material data, here taken from Refs.~\citenum{Green,RefractiveIndexInfo}. In order to use the material data to perform calculations within DGTD, we fit the measured data~\cite{Franta2017,RefractiveIndexInfo} to a 
Drude--Lorentz model~\cite{Si_lorentz2017,DrudeLorentz2024,PfeiferInterband2024},
\begin{equation} \label{eq:Drude_lorentz}
    \varepsilon (\omega)=\varepsilon_{\infty}-\frac{\omega_{\mathrm p}^2}{\omega^2+{\mathrm i}\gamma_{\mathrm D}\omega}+\sum_j \frac{f_j\omega_{0,j}^2}{\omega_{0,j}^2-\omega^2-{\mathrm i}\gamma_{\mathrm L,j}\omega}.
\end{equation}
Here, $\varepsilon_{\infty}$ denotes the background permittivity, $\omega_{\mathrm p}$ and $\gamma_{\mathrm D}$ represent, respectively, the plasma frequency and damping frequency in the Drude (second) term, while the $\omega_{0,j}$, and $\gamma_{\mathrm L,j}$ are the resonance frequencies and damping constants of the $j$th Lorentzian functions (last term); $f_j$ gives the strength for each Lorentz pole. 
Note, that we only introduce a limited number of Lorentz poles within the energy range of interest, while higher-energy interband transitions are included into the background permittivity $\varepsilon_{\infty}$.
The fitting parameters are shown in Sec.~S1 of the supplementary information (SI) (see also Refs.~\citenum{SiliconSpringer,Si2015gap,SiPhonon2012, SiWavelength1970,Green} therein).
Specifically, both the Drude term and a number of Lorentz poles are essential to achieve an optimum fitting result. Note that in the fitting process we constrain the parameters to be positive, for a realistic physical material model. Each Lorentz oscillator conceptually represents a bound--electron resonance, while the Drude term at lower energy is a pragmatic, yet fully causal, approach that enables the fitting procedure to capture a slowly varying background permittivity in this energy range, without implying any physical contribution from free carriers in Si.

With the material fitting data at hand, we perform DGTD simulations for swift electrons with different velocities penetrating a Si nanosphere of radius $R=79$\,nm, with impact parameter $b=0$. In the SI we also show the computational details and a consistency study for the CL spectra between DGTD and the analytical Mie solutions~\cite{Elli_Mie_code,Elli_Wenhua2023} carried out with both, fitted data and direct interpolation of the experimental data for Si.
There, we observe a very good agreement between the analytical and numerical results, thus justifying the Drude--Lorentz modeling in Eq.~\eqref{eq:Drude_lorentz}.

\subsection{Cathodoluminescence theory} \label{subsec: CL_theory}
As the swift electron beam traverses the dielectric NP, it excites polarization currents and geometric resonances that give rise to various radiation channels, including TR, CR, multipole Mie-type resonances and bulk-plasmon mediated emission~\cite{Dong2019}, part of which can couple to propagating modes and radiate to the far field, which can be analyzed as CL signal~\cite{ebel_nanoph14,Abajo2021,CLreview2023}. To characterize its time-domain radiation dynamics, we consider the time-dependent angle-resolved Poynting flux through a spherical surface in the far field with $\hat{\mathbf{r}}$ being the radial unit vector normal to it,
\begin{align} \label{eq:CL_time}
    \mathcal{P} (\theta, t) &= \frac{r^2}{\pi} \, 
    \, \int \mathrm{d}\phi \, \big\{ \mathbf{E}_\mathrm{ind} (\mathbf{r}, t) \times 
    \mathbf{H}_\mathrm{ind} (\mathbf{r}, t) \big\} \cdot \hat{\mathbf{r}},
\end{align}
which gives the angular radiated power in units of Watt~(W). Here, $\theta$ denotes the polar angle measured with respect to $z$-axis, as defined in Fig.~\ref{fig:Fig_1}, and $\phi$ represents the azimuthal angle in the $xy$-plane.
To characterize the frequency-domain radiation behavior, we evaluate the angular CL emission probability,
\begin{align} \label{eq:CL_omega}
    \Gamma_\mathrm{CL} (\theta,\omega) = \frac{er^2}{\pi \hbar^2\omega} \, \int \mathrm{d}\phi \, \mathrm{Re} \big\{ \mathbf{E}_\mathrm{ind} (\mathbf{r}, \omega) \times \mathbf{H}_\mathrm{ind}^* (\mathbf{r}, \omega) \big\} \cdot  \hat{\mathbf{r}},
\end{align}
with the fields $\mathbf{E}_\mathrm{ind}$ and $\mathbf{H}_\mathrm{ind}$ directly recorded from the DGTD computations. Note that $e$ is the elementary charge and $\Gamma_\mathrm{CL}$ has units of inverse energy ($\mathrm{eV}^{-1}$), which gives a spectral probability density per unit energy~\cite{GarciadeAbajo:1999prb,paper149}. 
Experimentally, angle-resolved CL is commonly measured by projecting the Fourier-plane emission onto a spectrograph slit, which selects a finite range of azimuthal angles $\phi$ while resolving polar-angle $\theta$ dependence~\cite{Polman2019_energy_momentum_CL}. Accordingly, the measured signal corresponds to an integration over the angular acceptance range rather than over the full azimuthal range used in Eq.~\eqref{eq:CL_omega}. However, for the central electron trajectory and spherical geometry considered here, the system is azimuthally symmetric. Therefore, a slit-selected slice changes only the overall CL intensity and does not change the polar-angle or spectral dependence.
By integrating Eq.~\eqref{eq:CL_omega} over the polar angle $\theta$ we obtain the total CL, which, in a Mie-based theory~\cite{Elli_Mie_code}, is evaluated in the far field ($k_0 r \to \infty$),
\begin{equation} \label{CL_decomposition}
       \Gamma_\mathrm{CL} (\omega) =\frac{1}{\pi \hbar\omega Z_0 k^2_0}  \sum_{\ell =1}^{\infty} \sum_{m=-\ell}^{+\ell}  \Big\{\big| b_{\ell  m} \big|^2 + \big| a_{\ell  m} \big|^2   \Big\},  
\end{equation}
with $a_{\ell m}$ and $b_{\ell m}$ being the Mie coefficients corresponding to each angular order, while $k_{0}$ is the wave number in vacuum, and $Z_{0}$ the corresponding impedance.

\subsection{Cherenkov and transition radiation}
\noindent \textbf{Cherenkov radiation.} The Vavilov--Cherenkov effect~\cite{TR_Ginzburg1996} describes the fact that electromagnetic (EM) radiation is emitted by a charged particle moving uniformly through a medium if the Cherenkov condition
\begin{equation} \label{eq:CR_condition}
    v_e > \frac{c}{n(\omega)}
\end{equation}
is satisfied, where $v_e=\beta c$ is the velocity of the charged particle and $0<\beta<1$ is the dimensionless measure of the relative velocity (see Sec.~S2 of the SI for further details). Consequently, when the velocity of the charged particle exceeds the phase velocity of light in the medium, Cherenkov radiation (CR) is emitted with a continuous spectrum and specific angular distribution. By neglecting the small energy of the photons we obtain the spectral condition of CR,
\begin{equation} \label{eq:CR_angle}
    \mathrm{cos}\alpha \approx \frac{1}{\beta n},
\end{equation}
where the angle $\alpha$ gives the emission angle of the radiated photons with respect to the propagation direction of the charged particle. We show in Sec.~S3 of the SI more details about the comparison between CR emission probability and the bulk loss in terms of analytical Mie-based EELS decomposition, which shows that CR indeed accounts for a large fraction of the electron's bulk loss.
However, due to material absorption/dispersion and boundary-induced reflections/refractions, most of the CR signals are trapped in the NP, or attenuated. For the material model with negligible loss, we expect part of this CR front to get transmitted through the boundary, which will then show fingerprints in the far-field CL/Poynting flux.

\noindent \textbf{Transition radiation.} When a charged particle penetrates the interface between two different media, TR~\cite{TR_Ginzburg1996,paper149,Jackson1998} can be excited as a consequence of the reorganization of the EM fields in different dielectric environments. This sudden change of the dielectric environments causes a polarization current that supports TR radiating into the far field~\cite{TR_Ginzburg1996,TR_Toth2004,TR_Chen2023,TR_Schmidt2018}. 
Specifically, at a vacuum-metal interface, the phenomenon can be interpreted through the formation of an induced "image" charge in the metal, which moves towards the electron~\cite{TR_Ginzburg1996}.
By crossing the boundary, the annihilation of both charges produces TR. For a general vacuum-dielectric interface, TR arises from the abrupt change in the EM field configuration required by the dielectric discontinuity. Nevertheless, the image-charge picture provides an intuitive interpretation of the mechanism underlying transition radiation. In contrast to CR, TR can, in principle, be excited at all electron velocities.

\section{Results and discussion} \label{sec:results}

In this section, we present the time- and frequency-domain results numerically via DGTD and analytically via Mie-based theory. First, we show the fingerprints of CR and TR for a dielectric sphere described by a constant-permittivity model, where we disentangle the contributions of CR and TR to the far-field radiation. Then we address the effect of different sphere sizes on TR, again for constant permittivity. Next, we consider the realistic Drude--Lorentz model for Si and demonstrate the near-to-far-field correspondence for the dipolar TR contributions. Finally, at high electron speeds far beyond CR threshold, we show the dominant role of TR for realistic Si in the far field via angle-resolved Poynting flux and CL emission probability studies.

\subsection{Silicon modeled via constant permittivity} \label{subsec:eps_16}

\subsubsection{Fingerprints of CR and TR}
%
\begin{figure*}[ht]
    \centering
    \includegraphics[trim = 8mm 40mm 5mm 30mm, clip, width=1.01\textwidth]{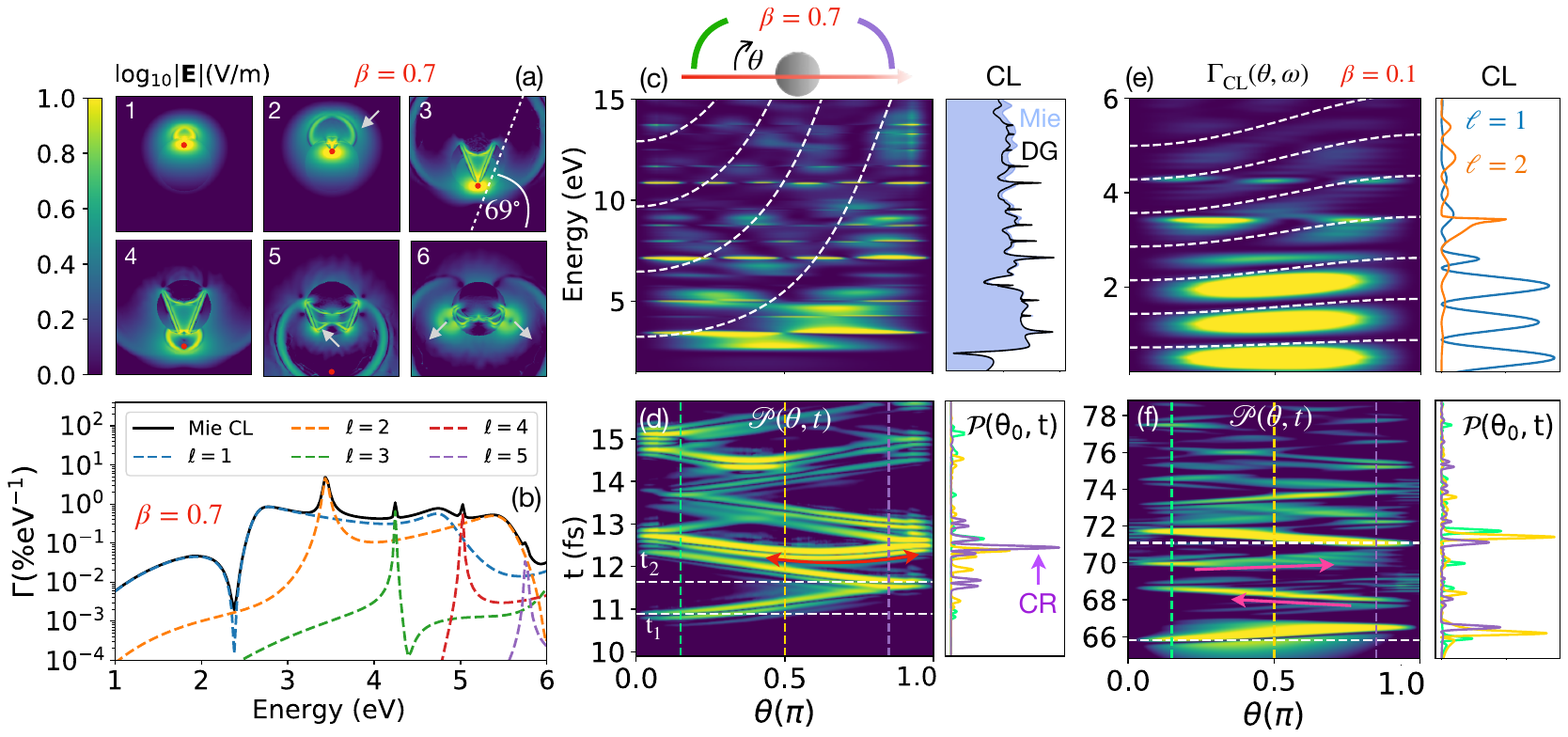}
    \caption{\textbf{Interaction between a swift electron beam and a dielectric sphere.} 
   We consider a free-standing dielectric sphere with $\varepsilon=16$ of radius $R=79$\,nm excited by a swift electron traveling downwards with velocity $\beta=0.7$, for penetrating trajectory $b=0$. 
   \textbf{\pnl{a}} Dynamics of the induced fields $|\mathbf{E}|$ for electron velocity $\beta=0.7$ at different times. 
   \textbf{\pnl{b}} Analytical multipole-decomposition of CL~\cite{Elli_Mie_code} in frequency-domain for a swift electron traveling with velocity $\beta=0.7$ at $b=1$\,nm. We only show the electric multipoles of order $\ell$ because higher-order magnetic modes $\ell>0$ are not excited for central electron trajectory.
   \textbf{\pnl{c}} Angle-resolved CL probability $\Gamma_\mathrm{CL}(\theta, \omega)$ for the angle $\theta$ (see inset) along the electron trajectory with the same setup as in \pnl{a}, where we observe the angular distribution of the sharp multipole resonances. 
   In the right inset, the total CL is plotted on a logarithmic scale as a black curve, where the Mie result from \pnl{b} is superimposed as a blue background.
   The interference condition in Eq.~\eqref{eq:double_slit_interference_condition} for $m=1,\ldots,4$ is superimposed onto the map as white-dashed curves. The definition of angles $\theta$ is shown in the upper inset.
   \textbf{\pnl{d}} Angle-resolved Poynting flux $\mathcal{P}(\theta, t)$ for the angle $\theta$ along the electron trajectory. The two dashed white lines give the arriving time of the two TR waves at angle $0$ and angle $\pi$. The red arrow indicates the escaped CR front that propagates into the far field. At fixed angles $\theta_0$ marked as vertical dashed lines, we plot $\mathcal{P}(\theta_0, t)$ on the right inset figure. There we identify the pulse peak for escaped CR radiation.
   \textbf{\pnl{e}} same as \pnl{c}, but for electron velocity $0.1c$. The interference condition in Eq.~\eqref{eq:double_slit_interference_condition} for $m=1,\ldots,7$ is superimposed onto the map as dashed white lines. We further plot the dipole (blue) and quadrupole (orange) components of total CL on the right inset.
   \textbf{\pnl{f}} same as \pnl{d}, but for electron velocity $\beta=0.1$. The pink arrows indicate secondary far-field radiation of TR.
    }
\label{fig:Fig_2}
\end{figure*}

As a first, simplistic model, we consider a levitated dielectric sphere with $\varepsilon=16+0^+\mathrm{i}$, where an infinitesimal imaginary part has been added for numerical stability in the Mie calculations~\cite{Elli_Mie_code}.
In this setup, the Si sphere with a radius of $R=79$\,nm is excited by a swift electron traveling downwards in the negative $z$ direction with velocity $\beta=0.7$, with a penetrating trajectory $b=0$, as illustrated in Fig.~\ref{fig:Fig_1}.
We note that the implicit assumption of a straight-line trajectory is well justified for electrons at $\beta \sim 0.7$ and the given Si particle size, based on Monte Carlo simulations of electron scattering in Si~\cite{Fiedler2021,ebel_nanoph14}. In contrast, experiments performed at $\beta\sim 0.14$ indicate significant beam broadening in 160\,nm-diameter Si spheres~\cite{Lebsir:2026}, a result that is also supported by Monte Carlo simulations.
More importantly, electron energy loss and the associated deceleration within the medium can cause significant deviations from the constant-velocity, no-recoil approximation, thus modifying the radiation pattern~\cite{Chahshouri_2022}.
As $\mathrm{Re}[\varepsilon]$ is constant and $\mathrm{Im}[\varepsilon] \sim 0$, no loss or dispersion is included, and the CR condition Eq.~\eqref{eq:CR_condition} with $\beta n(\omega)>1$ is strictly fulfilled. Via DGTD we solve Maxwell's equations and record the induced electric fields $\mathbf{E}$ at the $xz$-plane. The corresponding time-domain snapshots for the magnitude of the induced fields are shown in Fig.~\ref{fig:Fig_2}\pnl{a}, where the red dot denotes the position of the electron at each instance.
In Fig.~\ref{fig:Fig_2}\pnl{a-2}, shortly after the electron crosses the upper spherical boundary, we observe a clear front (see gray arrow) forming up and propagating into the vacuum, which is a result of the impulse kick from the electron at the entry point and the leaky Cherenkov front occurring at the same moment. After some time, in Fig.~\ref{fig:Fig_2}\pnl{a-3}, we observe a cone-shaped structure around the electron, indicating the presence of CR. By following Eq.~\eqref{eq:CR_angle} we confirm that the observed cone with angle $\alpha \approx \arccos(1/\beta n) \approx 69^\circ$ for $n=4$ is the Cherenkov front within the bulk. 
Further, as the electron crosses the exit point in Fig.~\ref{fig:Fig_2}\pnl{a-4}, we observe a similar propagating wave front emerging from the transition point. 
Shortly after that, in Fig.~\ref{fig:Fig_2}\pnl{a-5} we observe the part of the bulk CR front that is reflected back, indicated with a gray arrow. As the bulk front reaches the spherical boundary in Fig.~\ref{fig:Fig_2}\pnl{a-6}, we observe a transmitted/scattered spherical wave emerging from the cross points, indicated again as gray arrows. These describe the escaped CR front through the spherical boundary, which later contributes to the far-field radiation. For the full time-domain video via DGTD we refer to Sec.~S8 of the SI, where we also observe how the CR front is reflected back by the spherical boundary and contributes to the radiation in the backward direction. 
Our time-domain results therefore provide a dynamical complement to the existing frequency-domain framework~\cite{Abajo2004_Boundary_effects_CR}: we resolve the formation of TR at the two interfaces and the buildup of CR inside the sphere, thereby revealing how the finite boundary shapes the resulting emission.

In Fig.~\ref{fig:Fig_2}\pnl{b} we show the analytical multipole decomposition of the total CL, plotted in log-scale. Here, we consider an electron trajectory with $b=1$\,nm in Mie calculations to avoid divergence related to the source of EM waves passing through the origin. 
Considering a Gaussian charge distribution with a width of $5$\,nm in DGTD, $b=1$\,nm lies safely inside it, thus correctly reflecting the spectra for the central trajectory.
Each resonance peak can be identified as a specific multipole resonance. In this simplified material model, we can distinguish the multipole resonances from each other, which facilitates the understanding for each multipole contribution.
Note that we only show the electric multipoles because only the $m=0$ magnetic modes are excited due to the azimuthal symmetry of the system. To excite higher-order magnetic multipoles we need to break this symmetry, i.e., when the electron does not go through the center of the sphere.

Figure~\ref{fig:Fig_2}\pnl{c} shows the angle-resolved CL probability $\Gamma_\mathrm{CL}(\theta, \omega)$ in the frequency domain, directly computed from the fields calculated in the DGTD method. The definition of angle $\theta$ is shown in the inset. The total CL is shown in the right inset of the map, denoted by a black curve. We observe the clear angular distribution of different resonance peaks, which helps us to understand the nature of different excitations. 
Furthermore, we have superimposed the Mie result from Fig.~\ref{fig:Fig_2}\pnl{b} into the right inset of Fig.~\ref{fig:Fig_2}\pnl{c} as a blue background. 
For better comparison, both spectra are plotted using a logarithmic scale. 
The Mie-theory calculation uses a weakly absorptive permittivity, $\varepsilon=16+0.1\mathrm{i}$, and an impact parameter of $b=1$\,nm, whereas the DGTD calculation uses the lossless value $\varepsilon=16$ and $b=0$. 
The two methods predict consistent resonance positions, whereas at higher energies, the discrepancy is slightly larger, because for the mesh size and simulation time in the DGTD calculations considered here, we do not expect the results to converge at higher energies. 

To better understand the far-field behavior related to TR, we consider a simplified interference model, as shown in Fig.~\ref{fig:Fig_1}, where two light sources at transition points $z_1$ and $z_2$ start to propagate to the far field with a time delay $2R/v_e$, which gives the path time of the electron across the sphere. The detection distance is $R_\mathrm{CL}\sim 600$\,nm away, and the separation of the two emission points is $|z_2-z_1|=2R=158$\,nm, which satisfies $R_\mathrm{CL} > (2R)^2/\lambda$ for the studied energy range, thus justifying the Fraunhofer regime.
Therefore, the two light sources interfere like plane waves in the far field. Then the total far-field phase difference between two sources can be written as,
\begin{multline} \label{eq:TR_interference_phase_difference}
      \Delta\phi \approx \left[k_{\text{out}} \cos (\pi - \theta) -\frac{\omega}{v_e}\right](z_2-z_1)+ \Delta \phi_\ell(\omega)\\ \overset{!}{=} 2m\pi,
\end{multline}
which gives the constructive interference condition for integer number $m$. Here, $k_\mathrm{out} = \omega/c$ gives the free-space wavenumber for energy $\hbar \omega$, and $\theta$ corresponds to the angle relative to the $z$-axis, which is opposite to the moving direction of the electron.
Note that we add a mode specific phase term $\Delta \phi_\ell(\omega)$ that describes the scattering phase difference at two TR points for different multipole modes $\ell$. This term modifies the interference condition. However, to get a general idea of the "double-slit" interference effect, we neglect this term for the following study. The constructive interference condition gives the $\omega$--$\theta$ relation,
\begin{equation} \label{eq:double_slit_interference_condition}
    \omega(\theta)=\frac{c\,m\pi}{R}\frac{\beta}{1+\beta\cos\theta},
\end{equation}
for different orders $m$. In Fig.~\ref{fig:Fig_2}\pnl{c} we superimpose the condition of Eq.~\eqref{eq:double_slit_interference_condition} for $m=1,\ldots,4$ onto the angle-resolved CL map as white dashed lines. We observe only a very weak interference pattern in the map, that roughly corresponds to the $m=1$ order, because at such high electron velocities TR is not dominant in the CL. 

To understand the origin of this observation, we study the far-field photon emission dynamics in time. For this purpose, we plot in Fig.~\ref{fig:Fig_2}\pnl{d} the angle-resolved Poynting flux $\mathcal{P}(\theta, t)$ for the angle $\theta$ along the electron trajectory defined in Fig.~\ref{fig:Fig_2}\pnl{c}. 
As earlier shown in Fig.~\ref{fig:Fig_2}\pnl{a}, the TR is first induced at the transition points and then propagates with light velocity into the far field.
We thus define two reference time points: $t_1$ corresponds to the time point that the TR pulse emerging from the entry point reaches the detector, at angle $0$, and $t_2$ corresponds to the time point that the TR pulse emerging from the exit point reaches the detector, at angle $\pi$. We plot two guidelines (horizontal white dashed lines) in the map, corresponding to the time $t_1$ and $t_2$.
Starting from $t_1$, we observe a well-defined propagation of the waves, that reach further angles $0 \rightarrow \pi$ at slight delays. Then, starting from $t_2$, we observe the TR emerging from the exit point propagate into backwards angles $\pi \rightarrow 0$, which is clearly the evidence for TR radiation in the time-domain.
Further, as indicated in Fig.~\ref{fig:Fig_2}\pnl{a-6}, the Cherenkov front reaches the boundary and gets partially transmitted/scattered. These transmitted photons can be observed here as the bright fronts denoted by red arrows indicating the spreading direction of the front into different angles. 
This has important implications in applications, where the material properties could be inferred when the Cherenkov angle is determined in the far-field (the first arrival angle of the CR flow in the Poynting flux), particularly in transparent media.
The transmitted CR waves build a rather strong enhancement in the $\mathcal{P}(\theta, t)$ map, which indicates that CR is relatively strong in the far-field radiation, therefore TR does not dominate. 
At fixed angles $\theta_0$ marked as vertical dashed lines, we plot $\mathcal{P}(\theta_0, t)$ on the right inset of Fig.~\ref{fig:Fig_2}\pnl{d}. There, we identify the pulse peak associated with the escaped CR radiation, which is much stronger than TR for a forward angle denoted by violet color.
This further confirms the observation in Fig.~\ref{fig:Fig_2}\pnl{c}, where no clear TR interference pattern is observed. We thus conclude that the excitation of CR destroys the clear interference pattern of TR in a lossless dielectric sphere. Or rather: the strength of CR is compatible with TR.

As a comparison, we expand this study to lower electron velocity $\beta=0.1$, as shown in Figs.~\ref{fig:Fig_2}\pnl{e-f}, where no CR is excited. The angle-resolved CL map in Fig.~\ref{fig:Fig_2}\pnl{e} shows a clear TR interference pattern, where the condition in Eq.~\eqref{eq:double_slit_interference_condition} for $m=1,\ldots,7$ is superimposed onto the CL map as dashed white lines. We do not expect a perfect match between the dashed lines and the interference CL pattern, because the double-slit model is simplified without considering mode-specific phase terms. 
Further, on the right inset of Fig.~\ref{fig:Fig_2}\pnl{e}, we plot the analytical dipole (blue) and quadrupole (orange) CL decomposition. We note that the interference pattern at lower energies $<3$\,eV is dominated by the dipole component, and at higher energies higher-order multipoles take the leading role.

For the dielectric sphere with permittivity $\varepsilon=16$, a relatively large radius together with negligible material damping support spectrally sharp and well-defined higher-order Mie resonances. Specifically, the large sphere size allows for higher-angular momentum channels with $\ell>1$ in the Mie expansion.
In this regime, the radiative loss dominates over absorption; therefore, multipolar modes beyond the dipole mode survive the material internal damping and contribute to pronounced resonances with narrow linewidths in the CL spectrum.

Figure~\ref{fig:Fig_2}\pnl{f} shows the angle-resolved Poynting flux in time, for lower electron velocity $\beta=0.1$. Here, we observe similar behavior of the TR, with starting points $t_1$ and $t_2$ denoted by dashed white lines.
We further observe energy flows between the times $t_1$ and $t_2$, denoted by pink arrows. Indeed, these are the secondary TR emissions. As the TR at the entry point is excited, one part propagates into free space, which is the primary TR signal. The other part gets partially trapped in the sphere, with a portion of it being transmitted and propagating in the opposite direction.
This explains the first pink arrow. The second one, at later time, is related with further reflections of TR within the sphere, which acts as a resonator that traps TR within it. The TR at the exit point goes through similar reflections, and the secondary TR propagation can be observed after $t_2$.

\subsubsection{Disentangling the CR and TR contributions}

\begin{figure}[ht]
    \centering
    \includegraphics[trim = 77mm 66mm 86mm 43mm, clip, width=0.5\textwidth]{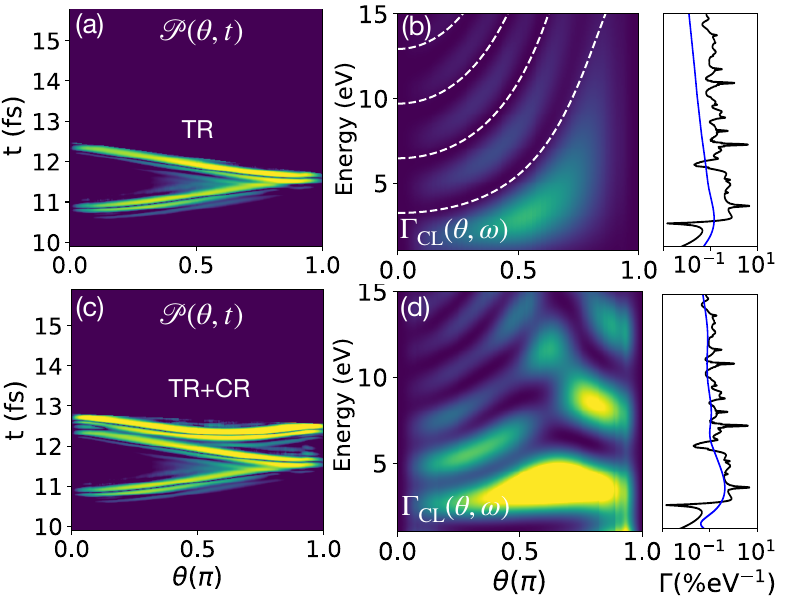}
    \caption{\textbf{Disentangling the CR and TR contributions for swift electron beams penetrating dielectric spheres.} 
   We consider a free-standing silicon sphere described by constant $\varepsilon=16$, excited by a swift electron with speed $\beta=0.7$ for penetrating trajectory $b=0$. By isolating the main TR and CR features from the Poynting flux in Fig.~\ref{fig:Fig_2}\pnl{d}, we calculate the corresponding CL according to Eq.~\eqref{eq:CL_omega}, thereby evaluating their roles in the far-field emission.
   \textbf{\pnl{a}} We apply a mask in terms of time $t$ and angle $\theta$ to isolate the main TR contributions in angle-resolved Poynting flux. 
   \textbf{\pnl{b}} We take the fields $\mathbf{E}$ and $\mathbf{H}$ corresponding to the mask in \pnl{a} and calculate the partial CL. The interference condition in Eq.~\eqref{eq:double_slit_interference_condition} for $m=1,\ldots,4$ is superimposed onto the map as white-dashed curves. By integrating the map over angle $\theta$ we obtain the corresponding CL spectrum, denoted by the blue curve in the right-hand sub-panel, while the black curve gives the total CL.
   \textbf{\pnl{c}} Similar as \pnl{a}, but we include the main CR contribution.
   \textbf{\pnl{d}} The angle-resolved CL corresponding to the mask in \pnl{c}.
    }
\label{fig:Fig_2_mask}
\end{figure}

As discussed in Fig.~\ref{fig:Fig_2}\pnl{c}, TR and CR play comparable roles in the far-field emission. Therefore, their interplay gives rise to the intricate features in the angle-resolved CL. To disentangle their roles, we start from the angle-resolved Poynting flux in Fig.~\ref{fig:Fig_2}\pnl{d} and isolate the main TR contribution by applying a mask in time $t$ and angle $\theta$, as shown in Fig.~\ref{fig:Fig_2_mask}\pnl{a}. Next, we take the fields $\mathbf{E}$ and $\mathbf{H}$ corresponding to the mask in \pnl{a} and calculate the angle-resolved CL according to Eq.~\eqref{eq:CL_omega}, as plotted in Fig.~\ref{fig:Fig_2_mask}\pnl{b}. We observe the characteristic double-slit interference patterns due to TR, as indicated by the white-dashed lines from the interference condition in Eq.~\eqref{eq:double_slit_interference_condition}. By integrating the angle-resolved CL map over angle $\theta$ we obtain the corresponding CL spectrum, denoted by the blue curve on the right side-panel of Fig.~\ref{fig:Fig_2_mask}\pnl{b}, while the black curve represents the total CL spectrum.
Then, we add the main CR contribution in the Poynting flux via appropriate mask, as shown in Fig.~\ref{fig:Fig_2_mask}\pnl{c}. The resulting CL is shown in Fig.~\ref{fig:Fig_2_mask}\pnl{d}, where we observe interference between the main TR and CR contributions. Indeed, these interference patterns build up the main features of the angle-resolved CL in Fig.~\ref{fig:Fig_2}\pnl{c}, in an envelope manner. In Sec.~S2 of the SI, we further present the study in which the CR part is isolated, revealing a smooth background in the CL. Interestingly, its interplay with TR shapes the angular distribution of the far-field radiation. When including all contributions of the Poynting flux, the complex CL map in Fig.~\ref{fig:Fig_2}\pnl{c} appears. 
Note that by isolating the main TR/CR burst we do not obtain sharp resonances in the CL map, for which we need to consider their contributions for longer times, including their reflections and secondary signals. 
In principle, by careful study of the time-domain dynamics, we are able to isolate the main TR and CR contributions; however, separating the individual contributions becomes increasingly difficult once they overlap, especially at longer times.
Therefore, it is important to combine both time- and frequency-domain analyses to identify the different radiation channels as much as possible.

\subsubsection{The effect of different sphere sizes}


\begin{figure}[ht]
    \centering
    \includegraphics[trim = 42mm 63mm 45mm 32mm, clip, width=0.5\textwidth]{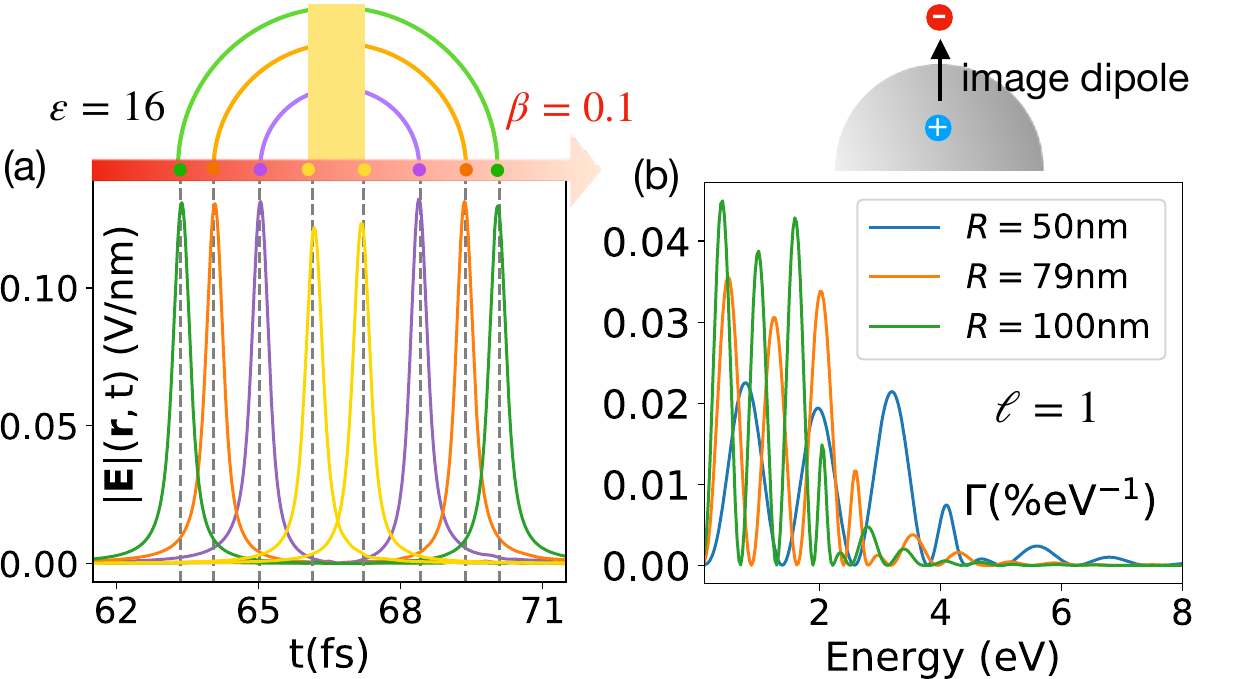}
    \caption{\textbf{Transition radiation at cross points for different sphere radii $R$.} 
   We consider a free-standing dielectric sphere described by constant $\varepsilon=16$, excited by a swift electron with speed $\beta=0.1$ for penetrating trajectory $b=0$. We consider different sphere radii $R=50, 79, 100$\,nm and a slab of thickness $d=30$\,nm.
   \textbf{\pnl{a}} Near-field components $\mathbf{E}(\mathbf{r},t)$ at the two transition points in time-domain. For different color (corresponding to different sphere sizes), the dashed gray lines denote the arriving time of the electron at the two transition points.
   \textbf{\pnl{b}} The dipole component $\ell=1$ of the analytical CL analysis, for different sphere sizes. In the upper inset we illustrate the induced image charge (blue dot with white plus sign) and the effective image dipole (black arrow), following Ref.~\citenum{paper149}.
    }
\label{fig:Fig_3}
\end{figure}

Previously, we have analyzed the CL mechanism for a dielectric sphere of radius $R=79$\,nm described by constant $\varepsilon=16$. Logically, it is interesting to know the effect of different sphere sizes. Therefore, we performed DGTD calculations for different sphere radii $R=50, 79, 100$\,nm at an electron impact distance $b=0$. The resulting near-field amplitudes $|\mathbf{E}(\mathbf{r},t)|$ at the two transition points in the time domain are shown in Fig.~\ref{fig:Fig_3}\pnl{a}, for different sphere sizes denoted by different colors. The gray-dashed lines correspond to the arriving time of the electron at the two transition points. For comparison, we also show the same study for a laterally infinitely large slab of thickness $d=30$\,nm.
We observe a pulse of induced fields $|\mathbf{E}(\mathbf{r},t)|$ as the electron crosses the TR points.
We further notice that the amplitude of the fields of smaller spheres is slightly stronger than the larger spheres, and the slab case shows the lowest pulse amplitudes. However, these differences are quite subtle. To explain this, we recall that the evanescent electric fields carried by a swift electron decay via modified Bessel functions with distance, through which the electron couples to the photonic modes in the NP. Around $2$\,eV the effective transverse field radius is a few nm, and for higher energies even smaller. The consequence is that the electron cannot distinguish much between different sphere sizes, because all of them are large enough so that they appear more or less flat for the impinging electron. But still, smaller spheres can concentrate light better, and therefore the field amplitude is slightly lower for larger spheres and the slab. On top of that, the width of the fields is related with the pulse width of the fields carried by the electron in the time-domain. A faster electron beam should create shorter pulse width compared with Fig.~\ref{fig:Fig_3}\pnl{a}.

Although the near-field dynamics at the transition points are more or less similar for different sphere sizes, the far-field radiation behavior could be quite different, as shown in Fig.~\ref{fig:Fig_3}\pnl{b}, where we show the dipole component $\ell=1$ of the analytical CL spectrum for different radii. We observe much stronger dipole radiation for larger spheres, which can be attributed to a larger polarization volume/charge separation in larger spheres if we treat the swift electrons (red dot with white minus sign) and the induced polarization charges (blue dot with white plus sign) in the NP as "image dipoles"~\cite{paper149}, denoted by a black arrow in the upper inset of Fig.~\ref{fig:Fig_3}\pnl{b}. Therefore, larger spheres support higher dipole moments and provide stronger dipole radiation in the far-field.

\subsection{Silicon modeled via realistic Drude--Lorentz model} \label{subsec:si_Drude_lorentz}

\subsubsection{Transition radiation from near-fields interference} \label{subsubsec:full_si_TR}
%
\begin{figure*}[ht]
    \centering
    \includegraphics[trim = 3mm 42mm 5mm 33mm, clip, width=1\textwidth]{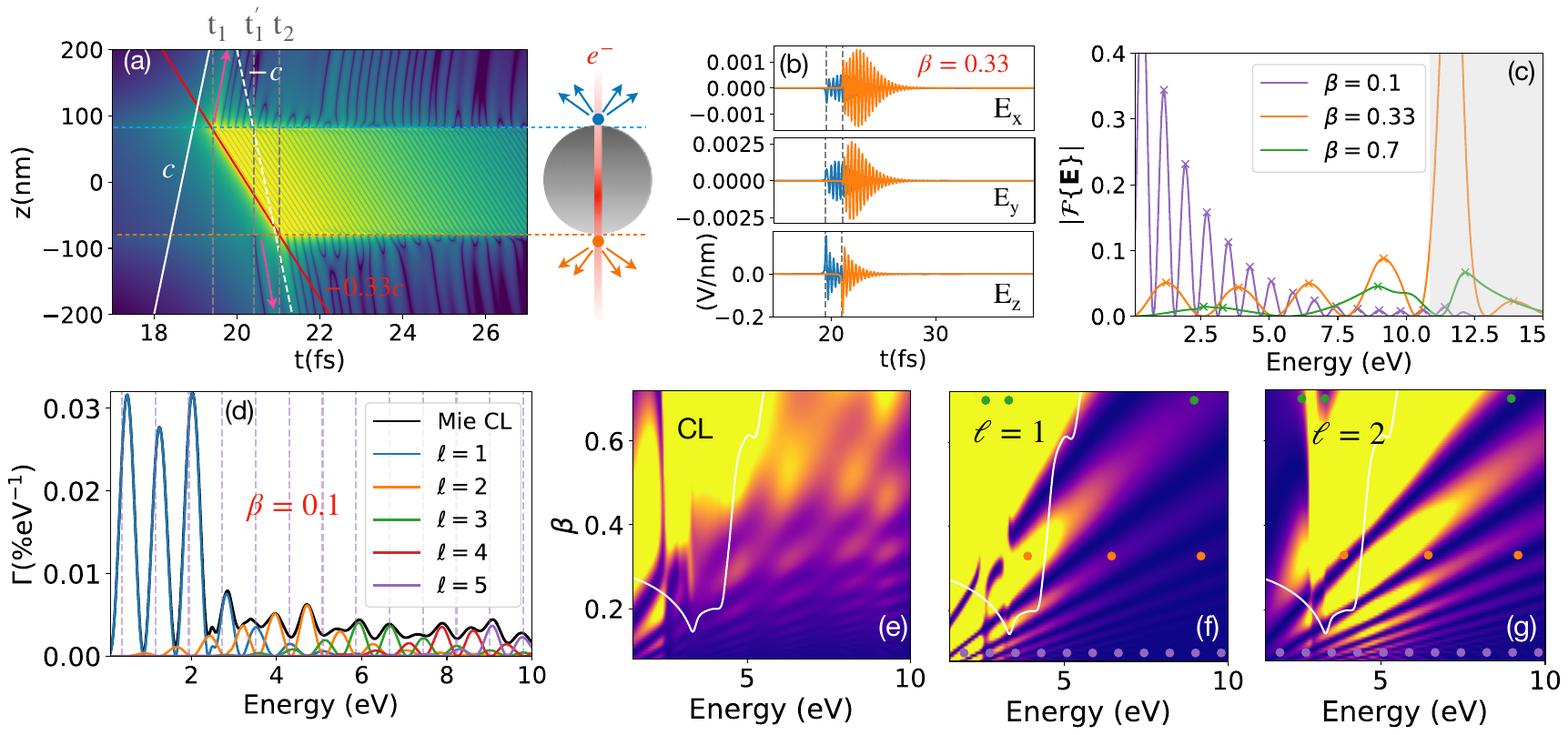}
    \caption{\textbf{Analytic and numerical TR for realistic Si modeling.} We consider a free-standing Si sphere of radius $R=79$\,nm described by the Drude--Lorentz modeling of Eq.~\eqref{eq:Drude_lorentz}, excited by a swift electron traveling downwards with different velocities for penetrating trajectory $b=0$. 
   \textbf{\pnl{a}}~Dynamics of the induced fields $|\mathbf{E}(z,t)|$ along the electron trajectory in time, for a swift electron traveling downwards with velocity $\beta=0.33$. The red line in \pnl{a} shows the electron trajectory in time and space. The dashed blue line indicates the upper boundary, and the dashed orange line indicates the lower boundary. 
   The dashed gray lines denote the three time points: $t_1$, when the electron crosses the entry point, $t_2$, when the electron crosses the exit point, and $t_1^{'}$, when the TR from the entry point arrives at the exit point.
   The setup is shown in the right inset. 
   \textbf{\pnl{b}} Near-field components $E_x(t)$, $E_y(t)$, and $E_z(t)$ at the two transition points, as line scans (blue and orange) from \pnl{a}. The dashed gray lines denote the arriving time of the electron at the two transition points.
   \textbf{\pnl{c}} Fourier transform of \pnl{b} into the frequency domain $|\mathcal{F} \{\mathbf{E}\}|$, with $\mathbf{E}=\mathbf{E}_1+\mathbf{E}_2$ the coherent sum of the fields at entry $\mathbf{E}_1$ and exit $\mathbf{E}_2$ points.
   \textbf{\pnl{d}} Analytical decomposition of the CL into different multipoles for $\beta=0.1$, with the peak positions in \pnl{c} superimposed as vertical dashed purple lines.
   \textbf{\pnl{e--g}} Analytical $\omega$--$\beta$ maps for full CL (left panel), electric-dipole component (middle panel), and electric-quadrupole component (right panel). The white solid line in each map gives the CR threshold according to Eq.~\eqref{eq:CR_condition}. The peak positions in \pnl{c} are superimposed onto the maps as dots for the electric-dipole $\ell =1$ and quadrupole $\ell =2$ contributions.
    }
\label{fig:Fig_4}
\end{figure*}

So far we have shown a thorough time- and frequency-dependent induced-field and CL study for a lossless and dispersion-free dielectric sphere, where TR and CR play comparable roles in the far-field radiation for high electron velocities. We now move on and consider a realistic Si material via Drude--Lorentz modeling in Eq.~\eqref{eq:Drude_lorentz}.
The additional Lorentz poles add absorption/dispersion to the material, compared to the constant real permittivity. Therefore, the time-domain dynamics and resonance features will be significantly modified. 

We start with the study of TR at the transition points as the electron crosses the spherical boundary. To understand the time dynamics at these points, we record the induced fields $|\mathbf{E}(t,z)|$ along the electron trajectory at different times, for a central trajectory $b=0$ and electron velocity $\beta \approx 0.33$, as shown in Fig.~\ref{fig:Fig_4}\pnl{a}.
The setup is shown in the right inset of Fig.~\ref{fig:Fig_4}\pnl{a}, where the swift electron travels downwards and penetrates the sphere at the blue (entry) and orange (exit) points. The red line in Fig.~\ref{fig:Fig_4}\pnl{a} shows the electron trajectory with $\beta \approx 0.33$ in time and space. The dashed blue line indicates the upper boundary and the dashed orange line indicates the lower boundary.
As the electron crosses the blue boundary, we observe light starting to propagate upwards at light velocity $c$, indicated as a white line. As the electron traverses the bulk, we observe phase-matched bulk polarization forming, indicated as parallel lines to the electron trajectory. As the electron exits the orange boundary, we further observe light starting to propagate downwards at light velocity, denoted by a dashed white line. The vertical dashed gray lines denote the three time points: $t_1$, when the electron crosses the entry point, $t_2$, when the electron crosses the exit point and $t_1^{'}$, when the TR from the entry point arrives at the exit point.

Next, we focus on the entry and exit points. For this we plot in Fig.~\ref{fig:Fig_4}\pnl{b} the three near-field components $E_x(t)$, $E_y(t)$, and $E_z(t)$, as line scans (blue and orange) from Fig.~\ref{fig:Fig_4}\pnl{a}. The dashed gray lines denote the arriving time of the electron at the two transition points. We observe strong delta-like impulsive fields at the transition points, with different phase relationship, followed by rather fast oscillations resulted from the bulk restoring force due to the material modeling.
In Sec.~S6 of the SI (see also Ref.~\citenum{TR_Chen2023} therein), we show an analytical study about the TR for swift electrons penetrating air-dielectric interface, where a delta-pulse in the time-domain is observed at transition point.
We note that the induced fields along the electron trajectory, $E_z$, are much stronger than the other two components. These fields correspond to the near fields excited by the electron, including TR and also evanescent fields due to bulk polarization.
We assume that parts of the fields in Fig.~\ref{fig:Fig_4}\pnl{b} propagate to the far field and build up a TR interference pattern, which could be observed in the CL. Note that the time-dependent fields at blue and orange transition points have a natural time delay, shown as the time difference between the dashed gray lines in Fig.~\ref{fig:Fig_4}\pnl{b}.
Therefore, we add these fields at two transition points coherently and Fourier-transform them into the frequency domain, as shown in Fig.~\ref{fig:Fig_4}\pnl{c}. There, we show the interference pattern for different electron velocities $\beta=0.1$, $\beta=0.33$, and $\beta=0.7$.
We observe the resulting peaks (colored crosses $\times$) due to the interference of two delta signals at specific time delays determined by the path time of the electron in the bulk~\cite{Fiedler2021}. At higher energy, around $>10$\,eV (gray background), we observe strong peaks as a consequence of the fast oscillations in Fig.~\ref{fig:Fig_4}\pnl{b}, which originate from the evanescent fields that do not propagate into the far field. 
Additionally in Sec.~S4 of the SI, we show similar studies for Si modeled with a constant permittivity by considering different electron speeds.

In Fig.~\ref{fig:Fig_4}\pnl{d} we show the analytical decomposition of the CL into different multipoles for $\beta=0.1$, with the peaks positions in Fig.~\ref{fig:Fig_4}\pnl{c} superimposed as vertical dashed purple lines.
We observe a good match between these lines and the electric-dipole peaks ($\ell =1$). We consider $\beta=0.1$ because at this electron velocity no CR is excited, which favors the observation of TR. Interestingly, the electric-quadrupole peaks are well phase-shifted to the dipole peaks, and we observe that the dashed lines correspond to the valleys of the quadrupole curve, thereby supporting the existence of distinct multipole scattering phases $\Delta \phi_l$ shown in Eq.~\eqref{eq:TR_interference_phase_difference}. We thus conclude that, for low velocities far below the Cherenkov threshold, the TR is responsible for the main resonance peaks of the dipole modes, as already shown in recent studies~\cite{Fiedler2021}. Remarkably, we can reproduce the far-field resonance positions in Fig.~\ref{fig:Fig_4}\pnl{d} via the interference pattern of the near-fields in Fig.~\ref{fig:Fig_4}\pnl{c}. 
We further notice that the resonances suffer from broadening due to absorption introduced by the Lorentz poles, compared to the sharp peaks for the constant $\varepsilon$ case, as shown in Fig.~\ref{fig:Fig_2}. As a matter of fact, higher-order multipoles $\ell >1$ are strongly damped and overlap with neighboring modes, see Fig.~\ref{fig:Fig_4}\pnl{d}.

To study the impact of different velocities, we show the analytical $\omega$--$\beta$ map for the full CL spectra, the electric-dipole contribution $\ell =1$, and the electric-quadrupole contribution $\ell =2$, in Figs.~\ref{fig:Fig_4}\pnl{e-g}. The white-solid line in each map gives the CR threshold according to Eq.~\eqref{eq:CR_condition}. We further superimpose the peaks from Fig.~\ref{fig:Fig_4}\pnl{c} onto the maps for the electric-dipole and quadrupole contributions, denoted by colored-dots in Fig.~\ref{fig:Fig_4}\pnl{f} and Fig.~\ref{fig:Fig_4}\pnl{g}, respectively. 
These points correspond to the interference maxima of the near-field dynamics at transition points for three different electron speeds denoted by different colors. 
The total CL map in Fig.~\ref{fig:Fig_4}\pnl{e} shows straight interference patterns at low energies corresponding to dipole interference, and curved patterns at higher energy governed by higher-order multipoles.
We can say that different multipoles govern different energy ranges. These interference patterns can be observed outside the CR threshold; Within it, CR smears the patterns. Furthermore, the peaks in Fig.~\ref{fig:Fig_4}\pnl{c} match the peaks of the dipolar component in Fig.~\ref{fig:Fig_4}\pnl{f}, and the valleys of the quadrupolar component in Fig.~\ref{fig:Fig_4}\pnl{g}. Therefore, we conclude that the near-field superposition at transition points governs the far-field radiation of the dipole component. At higher velocities ($\beta=0.7$) the behavior of the CL is rather complicated. To specify the main radiation channels for high electron velocity, we thus switch to a more detailed time-domain and frequency-domain study in the next section.

\subsubsection{Angle resolved study for full silicon model} \label{subsubsec:full_si_angle_resolved}

\begin{figure}[h!]
    \centering
    \includegraphics[trim = 83mm 43mm 70mm 40mm, clip, width=0.51\textwidth]{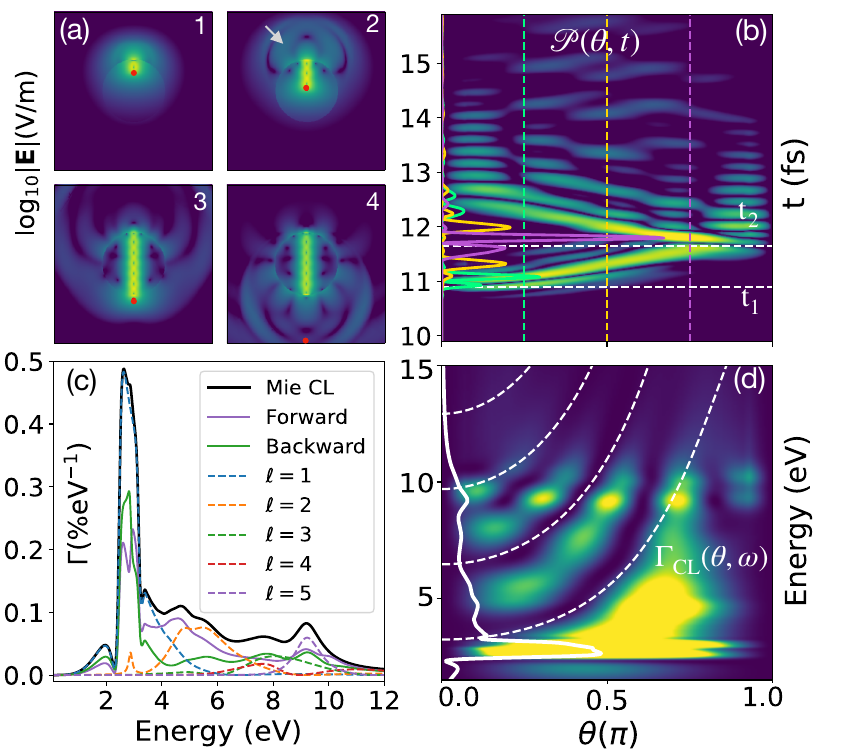}
    \caption{\textbf{Interaction between a swift electron beam and a Si sphere.}
   We consider a free-standing dielectric sphere of radius $R=79$\,nm, described by Drude--Lorentz modeling in Eq.~\eqref{eq:Drude_lorentz} with 8 Lorentz poles for Si material data in Ref.~\citenum{Franta2017}, excited by a swift electron traveling downwards with $\beta \approx 0.7$, for penetrating trajectory $b=0$.
   \textbf{\pnl{a}} Time-domain snapshots of the induced fields $|\mathbf{E}(\mathbf{r},t)|$ for different times $t$, plotted in log-scale.
   \textbf{\pnl{b}} Angle-resolved Poynting flux $\mathcal{P}(\theta, t)$ for the angle $\theta$ along the electron trajectory, as defined in Fig.~\ref{fig:Fig_1} and Fig.~\ref{fig:Fig_2}\pnl{c}.
   We further superimpose the temporal profiles $\mathcal{P}(\theta_0, t)$ for three selected angles $\theta_0$, indicated by dashed vertical lines, in the left corner of the map. 
   \textbf{\pnl{c}} Analytical multipole decomposition~\cite{Elli_Mie_code} of CL spectrum in frequency-domain, with $\ell$ denoting different electric multipole order. We also plot the total forward CL and backward CL via DGTD, by integrating \pnl{d} over forward ($\pi/2$ -- $\pi$) and backward (0 -- $\pi/2$) angles.
   \textbf{\pnl{d}}~Angle-resolved $\Gamma_\mathrm{CL}(\theta, \omega)$, where we observe the angular distribution of the multipole resonances. The total CL spectrum shown in \pnl{c} is superimposed in the left corner of the map as a solid-white line. Further, the constructive interference condition in Eq.~\eqref{eq:double_slit_interference_condition} for $m=1,\ldots,4$ is denoted by white-dashed lines. 
    }
\label{fig:Fig_5}
\end{figure}

To obtain further insight into the angular radiation behavior of different far-field photon emission channels when considering the full Drude--Lorentz modeling of Si, we perform 
DGTD calculations for high electron speed $\beta \approx 0.7$, where CR condition is satisfied. 
We first show time-domain snapshots for the induced electric fields $|\mathbf{E}(\mathbf{r},t)|$ at the $xz$-plane in Fig.~\ref{fig:Fig_5}\pnl{a}, for penetrating trajectory at impact distance $b=0$. Note that the material data are fitted up to 30\,eV to obtain a reliable time-domain behavior, see more details in Sec.~S1 of the SI. As the electron beam penetrates the first cross point, TR starts to form, shown as a strong field enhancement at the cross point in Fig.~\ref{fig:Fig_5}\pnl{a-1}. Later, in Fig.~\ref{fig:Fig_5}\pnl{a-2} we observe the TR front denoted by a gray arrow. 
As the electron travels inside the silicon sphere, we observe bulk polarization following the electron. We can also compare the effect shown here with Fig.~\ref{fig:Fig_2}, where a positive real constant $\varepsilon$ was considered and no bulk polarization was observed.
Indeed, the electron loses a significant amount of energy to excite bulk polarization in a realistic silicon sphere, including bulk plasmon excitations, see the strong bulk plasmon peak in EEL probability in Sec.~S3 of the SI. Although the electron velocity with $\beta=0.7$ is well beyond the Cherenkov threshold, Eq.~\eqref{eq:CR_condition}, for the low energy range, we did not observe any coherent CR front building up, in contrast to the material with constant permittivity shown in Fig.~\ref{fig:Fig_2}\pnl{a}. The reason is that in the realistic material model, we have to add several Lorentz poles to correctly model the high-energy behavior.
These Lorentz poles correspond to strong damping and absorption; therefore, CR is strongly suppressed in the high-energy range. In Sec.~S5 of the SI, we also show the field dynamics if we model Si with fewer Lorentz poles, where CR is less attenuated in high-energy range $>20$\,eV. Note that a correct material modeling up to higher energy range (above the bulk plasmon energy $\sim 16-17$\,eV~\cite{egerton2011electron}) is crucial to obtain reliable time-domain dynamics.

Next, to understand the far-field power flow in time, we plot in Fig.~\ref{fig:Fig_5}\pnl{b} the angle-resolved Poynting flux $\mathcal{P}(\theta, t)$ for the angle $\theta$ along the electron trajectory, as defined in Fig.~\ref{fig:Fig_1}. The angular flux $\mathcal{P}(\theta_0, t)$ at fixed angles $\theta_0$, indicated by the vertical-dashed lines, are superimposed in the left corner of the map. 
We add two guidelines (horizontal dashed white lines), which give the time $t_1$ and $t_2$, similar as in Fig.~\ref{fig:Fig_2}. Starting from $t_1$, we observe a well-defined propagation of the waves, that reach further angles at slight delays. Then, starting from $t_2$, we observe the TR emerging from the exit point propagate into backward angles, which is clearly the evidence for TR radiation in the time domain. When compared with Fig.~\ref{fig:Fig_2}\pnl{d} with constant permittivity, we notice that CR flux is not clearly visible here, thus justifying the dominant role of TR for a real silicon sphere at electron speed $\beta=0.7$.
For the forward angle, indicated by a purple curve, we observe a strong pulse in the Poynting flux in time, which is much stronger than for other angles. We assume that part of the CR waves escapes the spherical boundary and contributes to the far-field Poynting flux, although no clear CR front is observed in \pnl{a}. Such kind of strong energy flow in form of a pulse could be further studied for the generation of light pulse at specific direction. At smaller angles, the two peaks along each specific angle correspond to the TR that reaches the angle at different time delays.

In Fig.~\ref{fig:Fig_5}\pnl{c}, we perform an analytical decomposition of CL in frequency-domain to identify the contribution of each multipoles, where we also superimpose the total forward CL and backward CL, by integrating \pnl{d} over forward ($\pi/2$ -- $\pi$) and backward (0 -- $\pi/2$) angles. This helps us to determine the directional radiation characteristic of various multipoles.
Specifically, the peak splitting around 2.5–3\,eV in the forward direction should be attributed to the forward emission of the quadrupole, which appears due to large sphere retardation and is favored by the electron momentum direction, possibly also fed by weakly-radiated CR. Furthermore, the dipole channel exhibits dips around 2–4\,eV, which is a sign of destructive interference within dominant dipolar response due to retardation in large sphere, also encoded in the oscillatory Mie descriptions via Bessel functions $j_\ell$ and Hankel functions $h_\ell$, together with the phase accompanied with electron motion $e^{-\mathrm{i}\omega z/v_e}$~\cite{Elli_Mie_code,paper149}. We expect less such kind of effects in smaller spheres where quasi-static response applies.

The total CL spectrum in Fig.~\ref{fig:Fig_5}\pnl{c} tells us the total multipolar contributions integrated over solid angle $\Omega=(\theta,\phi)$. To further analyze the frequency-domain emission characteristics in greater detail, we plot in Fig.~\ref{fig:Fig_5}\pnl{d} the angle-resolved $\Gamma_\mathrm{CL}(\theta, \omega)$, where the integrated total CL spectrum is superimposed into the map on the left corner.
We observe clear curved branches on the map compared to Fig.~\ref{fig:Fig_2}\pnl{c} with constant permittivity, which indicates a strong interference effect associated with the TR, as supported by the pronounced TR-related power flow in Fig.~\ref{fig:Fig_5}\pnl{b}. Therefore, we superimpose the constructive interference condition in Eq.~\eqref{eq:double_slit_interference_condition} for $m=1,\ldots,4$ into the angle-resolved CL map as white-dashed lines. Together, Fig.~\ref{fig:Fig_5}\pnl{b} and Fig.~\ref{fig:Fig_5}\pnl{c} indicate that, TR dominates in the far-field CL signal for the material and electron speed under study.

Additionally, we investigate in Sec.~S7 of the SI the influence of different substrate materials on the resulting CL spectra, because in real CL experiments, the silicon sphere is usually deposited on a substrate~\cite{Lebsir:2026}. Among the different substrate materials, the influence of $\mathrm{Si_3N_4}$ is negligible, although weak guided modes, TR and CR (when permitted) could be excited in the substrate, but their amplitudes are comparatively weak as long as the substrate is thin enough. 
On the contrast, even very thin metallic substrates introduce undesired strong SPP excitations that overweigh the resonances in the Si, due to the strong plasmonic field enhancement supported by metallic structures.

\section{Conclusions}

We have presented a comprehensive combined time-domain/frequency-domain study for swift electron beams penetrating dielectric spheres using material models of different sophistication, including a non-dispersive material with constant permittivity and a Drude--Lorentz material model. Specifically, we have focused on the study of TR and CR of silicon spheres.
First, we have considered a lossless and dispersionless dielectric sphere modeled with a constant real permittivity. 
For an electron velocity over the Cherenkov threshold, we have confirmed the emergence of CR, which corresponds to the theoretical prediction of the Cherenkov angle. The time-domain Poynting flux shows clear evidence for TR that emerges from entry/exit points and spreads over further angles at later times. 
By isolating the main TR and CR features in Poynting flux, we have investigated their respective contributions to the CL probability. Indeed, their interplay shapes the main far-field radiation features.
As a result, we have found that CR and TR play comparable roles in the far-field radiation for the constant lossless permittivity. Further, we have demonstrated that when the material is modeled with a more realistic Drude--Lorentz model that accurately reproduces experimental data, the far-field CL -- even for high electron velocity above the CR threshold -- is dominated by TR. We have observed this TR signal as an interference between the fields originating from the entry and exit points. Consequently and interestingly, the near-field dynamics at the transition points governs the far-field dipole radiation pattern. 
The large sphere which we have studied in our work supports dipole and higher-order multipole excitations, and thus we have identified corresponding curved interference patterns both, in the numerical energy--angle CL pattern and in the analytical energy--velocity CL pattern. 
Beyond this, the angle-resolved CL and Poynting flux shows the utility for studying directional radiation and for distinguishing different radiation channels at different directions. For applications, the TR interference patterns observed in angle-resolved CL can be utilized to determine the size of the Si sphere for a known electron velocity, or vice versa and possibly even the exact shapes of arbitrary nanoparticles can be retrieved when scanning the electron beam position. Closely related applications include depth and trajectory sensing based on the resulting interference patterns.

\section{Methods} \label{sec:methods}

\subsection{Numerical simulation via DGTD} \label{subsec:DGTD_methods}

We employ the DGTD~\cite{DGTD} approach to simulate the electromagnetic problem of a NP excited by an electron beam, which combines a piecewise polynomial spatial interpolation on an unstructured tetrahedral mesh with a fourth-order Runge--Kutta time integrator providing an explicit solver for Maxwell's equations in the time-domain
\begin{subequations} \label{Eq:Maxwell}
\begin{align}     
    \partial_t \mathbf{H} (\mathbf{r}, t) = &   
    - \mu^{-1}_0\mu^{-1}(\mathbf{r}) \,  \mathbf{\nabla} \times \mathbf{E} (\mathbf{r}, t),
    \\
    \partial_t \mathbf{E} (\mathbf{r}, t)=&  
    \, \varepsilon^{-1}_0\varepsilon^{-1}(\mathbf{r}) \, \left[ 
    \mathbf{\nabla} \times\mathbf{H} (\mathbf{r}, t) - \mathbf{j}(\mathbf{r}, t)\right].
\end{align}
\end{subequations}
Here, $\mathbf{j}$ is the total current density that encompasses both any current associated with the excitation source, as well as dispersive polarization currents~\cite{DGTD}. The resulting method is memory-efficient compared to traditional finite elements and especially well-suited for the calculation of wide-band spectra. With the unstructured tetrahedral mesh, it is possible to discretize the computational domain in different mesh sizes for different regions~\cite{geuzaine2009gmsh}, depending on whether the mesh size is small enough to resolve the rapid variation of the electromagnetic fields (see SI for further technical details). 

{\bf Electron source.} In particular, the electron source is implemented as a Gaussian charge distribution~\cite{Talebi_2013} of the form
\begin{equation}   \label{eq:Gaussiancharge}
    \rho(\mathbf{r}) = - \frac{ e}{\sigma_e^3 \sqrt{\pi^3} } \exp(-r^2/\sigma_e^2),
\end{equation}
with a width $\sigma_e=5$\,nm. This choice essentially prevents numerical artifacts arising when implementing a point-charge particle moving inside the NP. Specifically, the full relativistic, time-dependent incident fields carried by the swift electron in the reference frame are obtained via a Lorentz transformation of the fields from the electron's rest frame~\cite{Maciel_Efield_electron2019}.
As a time-domain solver, we can record the induced electromagnetic fields $\mathbf{E}^\mathrm{ind}(\mathbf{r},t)$ and $\mathbf{H}^\mathrm{ind}(\mathbf{r},t)$ both in the time-domain and subsequently in the frequency-domain by performing an on-the-fly Fourier transform within DGTD, with which we then compute angle-resolved Poynting flux $\mathcal{P}(\theta, t)$ and angle-resolved cathodoluminescence $\Gamma_\mathrm{CL}(\theta, \omega)$ of the electron when penetrating the NP following Eq.~\eqref{eq:CL_time} and Eq.~\eqref{eq:CL_omega}. By integrating Eq.~\eqref{eq:CL_omega} over angle $\theta$ we obtain the total CL spectrum. 

{\bf No-recoil approximation.} The total energy loss of the electron to the excitations in the NP is estimated to be $\Delta E \sim 1$\,eV, which is negligible compared to the electrons' kinetic energy (keV range). Additionally, the total momentum transfer from the electron to the nanowire is $\hbar k  \ll m_ev_e$, with $m_e$ the mass of the electron.
Therefore, we adopt the no-recoil approximation and assume a straight-line trajectory with constant velocity $v_e \hat{\mathbf{z}}^-=\beta c \hat{\mathbf{z}}^-$ in our numerical computations~\cite{paper149}, with $\hat{\mathbf{z}}^-$ denoting the negative $z$ direction.

\subsection{Analytic approach} \label{subsec:methods_analytics}

The analytical Mie-based calculations/multipolar-decompositions follow Ref.~\citenum{Elli_Mie_code} for the electron trajectory $b=1$\,nm due to the numerical stability. 
The electron is modeled via a point charge and the maximal multipole order is set to 5, which determines the multipole decomposition order $\ell$ shown in the main text.
For the material data, we considered constant permittivity $\varepsilon=16+0^+\mathrm{i}$, Si modeled via Drude--Lorentz fitting and interpolated experimental data according to Ref.~\citenum{Franta2017}.

Following Refs.~\citenum{Elli_Wenhua2023,Elli_Mie_code}, the electron is represented as a point charge $-e$ moving at a constant velocity $v$ along a straight trajectory parallel to the $z$ axis, $\mathbf{r}_{\mathrm{e}}(t)= b\hat{\mathbf{x}} + vt\hat{\mathbf{z}}$, within the no-recoil approximation. For $b<R$, the electron enters and exits the sphere at $z=\mp z_{\mathrm{e}}$, respectively, where $z_{\mathrm{e}}=\sqrt{R^{2}-b^{2}}$. The trajectory is therefore divided into an internal segment, $-z_{\mathrm{e}}<z<z_{\mathrm{e}}$, and two external segments. The direct electron field is calculated separately in these regions using the corresponding dielectric properties: the external contribution is constructed from the electron field in vacuum, whereas the internal contribution is obtained for an electron moving through a homogeneous medium with the permittivity of the sphere. These fields are expanded in spherical waves and combined with the boundary-induced fields by imposing continuity of the tangential electric and magnetic fields at $r=R$. The resulting exterior multipole coefficients determine the CL emission.

\section{Supporting Information}

Silicon material models; DGTD computational details; decomposition of the silicon-sphere EEL spectrum; substrate effects on the CL spectrum; TR at an dielectric-vacuum interface; and a time-domain movie of an electron traversing a dielectric sphere above the Cherenkov threshold.

\section{Acknowledgments}

%
The Center for Polariton-driven Light--Matter Interactions (POLIMA) is funded by the Danish National Research Foundation (Project No. DNRF165, POLIMA).
C.~T. acknowledges support from the Independent Research Fund Denmark (Grant No. 5281-00155B).
We thank P.~Elli~Stamatopoulou for providing the updated Mie-code.

\hfill

\bibliographystyle{achemso}
\bibliography{refs}

@article{paper149,
   author    = {F. J. {Garc\'{\i}a de Abajo}},
   title     = {Optical excitations in electron microscopy},
   journal   = {Reviews of Modern Physics},
   year      = 2010,
   volume    = 82,
   pages     = {209--275},
   doi       = {10.1103/RevModPhys.82.209}
}

@article{Polman_2019,
   title={Electron-beam spectroscopy for nanophotonics},
   volume={18},
   ISSN={1476-4660},
   url={http://dx.doi.org/10.1038/s41563-019-0409-1},
   DOI={10.1038/s41563-019-0409-1},
   number={11},
   journal={Nature Materials},
   publisher={Springer Science and Business Media LLC},
   author={Polman, Albert and Kociak, Mathieu and García de Abajo, F. Javier},
   year={2019},
   month=jul, pages={1158–1171} }

@article{talebi2017interaction,
  title={Interaction of electron beams with optical nanostructures and metamaterials: from coherent photon sources towards shaping the wave function},
  author={Talebi, Nahid},
  journal={Journal of Optics},
  volume={19},
  number={10},
  pages={103001},
  year={2017},
  publisher={IOP Publishing},
doi={10.1088/2040-8986/aa8041}
}

@article{Abajo2021,
author = {García de Abajo, F. Javier and Di Giulio, Valerio},
title = {Optical Excitations with Electron Beams: Challenges and Opportunities},
journal = {ACS Photonics},
volume = {8},
number = {4},
pages = {945-974},
year = {2021},
doi = {10.1021/acsphotonics.0c01950},
}

@article{CLreview2023,
author = {Dang, Zhibo and Chen, Yuxiang and Fang, Zheyu},
title = {Cathodoluminescence Nanoscopy: State of the Art and Beyond},
journal = {ACS Nano},
volume = {17},
number = {24},
pages = {24431-24448},
year = {2023},
doi = {10.1021/acsnano.3c07593}
}

@article{paper369,
Title={Can copper nanostructures sustain high-quality plasmons?},
author={V. Mkhitaryan and K. March and E. Tseng and X. Li and L. Scarabelli and L. M. Liz-Marz\'an and S.-Y. Chen and L. H. G. Tizei and O. St\'ephan and J.-M. Song and M. Kociak and F. J. {Garc\'{\i}a de Abajo} and A. Gloter},
journal={Nano Letters},
volume={21},
pages={2444--2452},
year={2021},
doi={10.1021/acs.nanolett.0c04667}
}

@article{hohenester2009electron,
  title = {Electron-Energy-Loss Spectra of Plasmonic Nanoparticles},
  author = {Hohenester, Ulrich and Ditlbacher, Harald and Krenn, Joachim R.},
  journal = {Physical Review Letters},
  volume = {103},
  issue = {10},
  pages = {106801},
  numpages = {4},
  year = {2009},
  month = {08},
  publisher = {American Physical Society},
  doi = {10.1103/PhysRevLett.103.106801},
  url = {https://link.aps.org/doi/10.1103/PhysRevLett.103.106801}
}

@article{paper085,
   author    = {J. Nelayah and M. Kociak and {O. St\'{e}phan} and F. J. {Garc\'{\i}a de Abajo}
                and M. Tenc\'e and L. Henrard and D. Taverna and I. Pastoriza-Santos
                and L. M. Liz-Marz\'{a}n and C. Colliex},
   title     = {Mapping surface plasmons on a single metallic nanoparticle},
   journal   = {Nature Physics},
   year      = 2007,
   volume    = 3,
   pages     = {348--353},
   doi       = {10.1038/nphys575}
}

@article{TR_Ginzburg1996,
doi = {10.1070/PU1996v039n10ABEH000171},
url = {https://doi.org/10.1070/PU1996v039n10ABEH000171},
year = {1996},
month = {oct},
publisher = {},
volume = {39},
number = {10},
pages = {973},
author = {Vitalii L Ginzburg},
title = {Radiation by uniformly moving sources ({Vavilov–Cherenkov} effect, transition radiation, and other phenomena)},
journal = {Physics-Uspekhi}
}

@article{TR_Chen2023,
author = {Ruoxi Chen  and Jialin Chen  and Zheng Gong  and Xinyan Zhang  and Xingjian Zhu  and Yi Yang  and Ido Kaminer  and Hongsheng Chen  and Baile Zhang  and Xiao Lin },
title = {Free-electron {Brewster}-transition radiation},
journal = {Science Advances},
volume = {9},
number = {32},
pages = {eadh8098},
year = {2023},
doi = {10.1126/sciadv.adh8098}
}

@article{Analytics_CR_TR_2003,
author = {Afanasiev, G. and Kartavenko, Vladimir and Stepanovskiy, Yuriy},
year = {2003},
month = {09},
pages = {4026–4056},
title = {{Vavilov–Cherenkov} and transition radiations on the dielectric and metallic spheres},
volume = {44},
journal = {Journal of Mathematical Physics},
doi = {10.1063/1.1602162}
}

@book{SiliconSpringer,
author={Safa Kasap and Peter Capper},
publisher = {Springer},
address = {Cham},
isbn = {978-3-319-48933-9},
title = {Springer Handbook of Electronic and Photonic Materials},
doi = {10.1007/978-3-319-48933-9},
year = {2017},
}

@inbook{Si_opticalProperties2025,
author = {Popescu, Dana Georgeta},
title = {Optical Properties of Silicon and Fundamentals of Waveguide Theory in Silicon Photonics},
booktitle = {Advanced Optoelectronics and Photonic Technologies- Fundamentals, Devices and Renewable Energy Applications},
publisher = {IntechOpen},
address = {London},
year = {2025},
editor = { Popescu, Dana Georgeta and  Nazir, Muhammad Shahzad},
chapter = {2},
doi = {10.5772/intechopen.1010928}
}

@article{Si_Tarrio1993,
author = {C. Tarrio and S. E. Schnatterly},
journal = {Journal of the Optical Society of America B},
number = {5},
pages = {952--957},
publisher = {Optica Publishing Group},
title = {Optical properties of silicon and its oxides},
volume = {10},
month = {May},
year = {1993},
url = {https://opg.optica.org/josab/abstract.cfm?URI=josab-10-5-952},
doi = {10.1364/JOSAB.10.000952},
}

@article{Maciel_Efield_electron2019,
author = {Maciel Escudero, Carlos and Reyes-Coronado, A.},
year = {2019},
month = {03},
pages = {137-149},
title = {Electromagnetic fields produced by a swift electron: A source of white light},
volume = {86},
journal = {Wave Motion},
doi = {10.1016/j.wavemoti.2019.01.005}
}

@article{Ritchie1988,
author = {R. H. Ritchie and A. Howie},
title = {Inelastic scattering probabilities in scanning transmission electron microscopy},
journal = {Philosophical Magazine A},
volume = {58},
number = {5},
pages = {753-767},
year = {1988},
doi = {10.1080/01418618808209951}
}

@article{Pogorzelski1973,
  title = {Diffraction Radiation from a Charged Particle Moving through a Penetrable Sphere},
  author = {Pogorzelski, R. and Yeh, C.},
  journal = {Physical Review A},
  volume = {8},
  issue = {1},
  pages = {137-144},
  numpages = {0},
  year = {1973},
  doi = {10.1103/PhysRevA.8.137}
}

@book{Jackson1998,
    author = {Jackson, John David},
    title = {Classical Electrodynamics},
    isbn = {978-0-471-30932-1},
    publisher = {Wiley \& Sons},
    address = {New York},
    year = {1998}
}

@article{Dong2019,
author = {Dong, Zhaogang and Wang, Tao and Chi, Xiao and Ho, Jinfa and Tserkezis, Christos and Yap, Sherry Lee Koon and Rusydi, Andrivo and Tjiptoharsono, Febiana and Thian, Dickson and Mortensen, N. Asger and Yang, Joel K. W.},
title = {Ultraviolet Interband Plasmonics With {Si} Nanostructures},
journal = {Nano Letters},
volume = {19},
number = {11},
pages = {8040-8048},
year = {2019},
doi = {10.1021/acs.nanolett.9b03243}
}

@article{PfeiferInterband2024,
author = {Pfeifer, Max and Huynh, Dan-Nha and Wegner, Gino and Intravaia, Francesco and Peschel, Ulf and Busch, Kurt},
year = {2024},
pages = {7},
title = {Time-domain modeling of interband transitions in plasmonic systems},
volume = {130},
journal = {Applied Physics B},
doi = {10.1007/s00340-023-08138-0}
}

@article{TserkezisHybridisation,
url = {https://doi.org/10.1515/nanoph-2023-0781},
title = {Self-hybridisation between interband transitions and {Mie} modes in dielectric nanoparticles},
author = {Tserkezis, Christos  and  Stamatopoulou, P. Elli and  Wolff, Christian and  Mortensen, N. Asger},
pages = {2513--2522},
volume = {13},
number = {14},
journal = {Nanophotonics},
doi = {10.1515/nanoph-2023-0781},
year = {2024},
lastchecked = {2025-11-14}
}

@article{Fiedler2021,
author = {Fiedler, Saskia and Stamatopoulou, P. and Assadillayev, Artyom and Wolff, Christian and Sugimoto, Hiroshi and Fujii, Minoru and Mortensen, N. A. and Raza, Søren and Tserkezis, Christos},
year = {2021},
month = {12},
volume = {22},
pages = {2320-2327},
journal = {Nano Letters},
title = {Disentangling cathodoluminescence spectra in nanophotonics: particle eigenmodes \emph{vs} transition radiation},
doi = {10.1021/acs.nanolett.1c04754}
}

@article{ebel_nanoph14,
author = {Ebel, S. and Lebsir, Y. and Yezekyan, T. and
Mortensen, N.~A. and Morozov, S.},
title = {An atlas of photonic and plasmonic materials
for cathodoluminescence microscopy},
journal = {Nanophotonics},
volume = {14},
number = {15},
pages = {2647--2671},
year = {2024},
doi = {10.1515/nanoph-2025-0135}
}

@INPROCEEDINGS{Si_lorentz2017,
  author={Sehmi, H. S. and Langbein, W. and Muljarov, E. A.},
  booktitle={2017 Progress In Electromagnetics Research Symposium - Spring (PIERS)}, 
  title={Optimizing the {Drude--Lorentz} model for material permittivity: Examples for semiconductors}, 
  year={2017},
  volume={},
  number={},
  pages={994-1000},
  doi={10.1109/PIERS.2017.8261889}}

@article{DrudeLorentz2024,
author = {Ben Soltane, Isam and Dierick, Félice and Stout, Brian and Bonod, Nicolas},
title = {Generalized {Drude–Lorentz} Model Complying with the Singularity Expansion Method},
journal = {Advanced Optical Materials},
volume = {12},
number = {12},
pages = {2400093},
doi = {10.1002/adom.202400093},
year = {2024}
}

@Article{DGTD,
  Title                    = {Discontinuous {Galerkin} methods in nanophotonics},
  Author                   = {K. Busch and M. K\"onig and J. Niegemann},
  year = {2011},
  month = {11},
  pages = {773-809},
  Volume                   = {5},
journal = {Laser \& Photonics Reviews},
doi = {10.1002/lpor.201000045}
}

@article{Elli_Wenhua2023,
       title = {Electron beams traversing spherical nanoparticles: Analytic and numerical treatment},
  author = {Stamatopoulou, P. Elli and Zhao, Wenhua and Rodr\'{\i}guez Echarri, \'Alvaro and Mortensen, N. Asger and Busch, Kurt and Tserkezis, Christos and Wolff, Christian},
  journal = {Physical Review Research},
  volume = {6},
  issue = {1},
  pages = {013239},
  numpages = {15},
  year = {2024},
  month = {03},
  publisher = {American Physical Society},
  doi = {10.1103/PhysRevResearch.6.013239}
}

@article{Kiel_metalspitze,
  title = {Real-space imaging of nanotip plasmons using electron energy loss spectroscopy},
  author = {Schr\"oder, Benjamin and Weber, Thorsten and Yalunin, Sergey V. and Kiel, Thomas and Matyssek, Christian and Sivis, Murat and Sch\"afer, Sascha and von Cube, Felix and Irsen, Stephan and Busch, Kurt and Ropers, Claus and Linden, Stefan},
  journal = {Physical Review B},
  volume = {92},
  issue = {8},
  pages = {085411},
  numpages = {7},
  year = {2015},
  month = {08},
  publisher = {American Physical Society},
  doi = {10.1103/PhysRevB.92.085411},
  url = {https://link.aps.org/doi/10.1103/PhysRevB.92.085411}
}

@article{Agwire2026,
  title = {Time-domain study of surface plasmon polariton propagation in silver nanowires},
  author = {Zhao, Wenhua and Echarri, \'Alvaro Rodr\'{\i}guez and Eljarrat, Alberto and Nerl, Hannah C. and Kiel, Thomas and Haas, Benedikt and Halim, Henry and Lu, Yan and Koch, Christoph T. and Busch, Kurt},
  journal = {Physical Review B},
  volume = {113},
  issue = {8},
  pages = {085425},
  numpages = {12},
  year = {2026},
  month = {Feb},
  publisher = {American Physical Society},
  doi = {10.1103/gpqj-4v29},
  url = {https://link.aps.org/doi/10.1103/gpqj-4v29}
}

@article{Green,
author = {Green, Martin},
year = {2008},
month = {11},
pages = {1305-1310},
title = {Self-consistent optical parameters of intrinsic silicon at {300 K} including temperature coefficients},
volume = {92},
journal = {Solar Energy Materials and Solar Cells},
doi = {10.1016/j.solmat.2008.06.009}
}

@article{RefractiveIndexInfo,
author = {Polyanskiy, Mikhail},
year = {2024},
title = {Refractiveindex.info database of optical constants},
volume = {11},
pages = {94},
journal = {Scientific Data},
doi = {10.1038/s41597-023-02898-2}
}

@article{Franta2017,
author = {Franta, Daniel and Dubroka, Adam and Wang, Chennan and Giglia, Angelo and Voh\'{a}nka, Ji\v{r}\'{i} and Franta, Pavel and Ohlı\'{i}dal, Ivan},
year = {2017},
month = {02},
pages = {405-419},
title = {Temperature-dependent dispersion model of float zone crystalline silicon},
volume = {421},
journal = {Applied Surface Science},
doi = {10.1016/j.apsusc.2017.02.021}
}

@article{Elli_Mie_code,
author = {P. Elli Stamatopoulou and Carlos Maciel-Escudero and N. Asger Mortensen and Carsten Rockstuhl and Christos Tserkezis},
journal = {Journal of the Optical Society of America B},
number = {7},
pages = {1620--1637},
publisher = {Optica Publishing Group},
title = {Analytic methods in electron energy-loss and cathodoluminescence spectroscopy of planar and spherical nanostructures: tutorial},
volume = {42},
month = {Jul},
year = {2025},
url = {https://opg.optica.org/josab/abstract.cfm?URI=josab-42-7-1620},
doi = {10.1364/JOSAB.559215}
}

@article{GarciadeAbajo:1999prb,
  title = {Relativistic energy loss and induced photon emission in the interaction of a dielectric sphere with an external electron beam},
  author = {Garc\'{i}a de Abajo, F. J.},
  journal = {Physical Review B},
  volume = {59},
  issue = {4},
  pages = {3095--3107},
  year = {1999},
  doi = {10.1103/PhysRevB.59.3095}
}

@article{TR_Toth2004,
author = {Toth, Csaba and Tilborg, Jeroen and Geddes, C and Fubiani, Gwenael and Schroeder, Carl and Esarey, Eric and Faure, Jerome and Dugan, G. and Leemans, Wim},
year = {2004},
month = {09},
pages = {491},
title = {Powerful pulsed {THz} radiation from laser-accelerated relativistic electron bunches},
volume = {5448},
journal = {Proceedings of SPIE},
doi = {10.1117/12.548945}
}

@article{Polman_CL_2016,
author = {Osorio, Clara I. and Coenen, Toon and Brenny, Benjamin J. M. and Polman, Albert and Koenderink, A. Femius},
title = {Angle-Resolved Cathodoluminescence Imaging Polarimetry},
journal = {ACS Photonics},
volume = {3},
number = {1},
pages = {147-154},
year = {2016},
doi = {10.1021/acsphotonics.5b00596}
}

@article{TR_Schmidt2018,
    author       = {Schmidt, Bernhard and Wesch, Stephan and Kövener, Toke and Behrens, Christopher and Hass, Eugen and Casalbuoni, Sara and Schmüser, Peter},
    year         = {2018},
    title        = {Longitudinal Bunch Diagnostics using Coherent Transition Radiation Spectroscopy},
    doi          = {10.5445/IR/1000088035},
    journal       = {DESY Reports},
    volume       = {18},
    number       = {027}
}

@book{egerton2011electron,
  title={Electron energy-loss spectroscopy in the electron microscope},
  author={Egerton, Ray F},
  publisher = {Springer},
  address = {New York},
   year      = 2011,
   OPTpages={XII, 491},
   doi={10.1007/978-1-4419-9583-4}
}

@article{Lebsir:2026,
      title={Probe- and Substrate-Dependent Visibility of {Mie} Resonances in Silicon Nanospheres}, 
      author={Yonas Lebsir and Huatian Hu and P. A. D. Gonçalves and Hiroshi Sugimoto and Minoru Fujii and N. Asger Mortensen and Christos Tserkezis and Sergii Morozov},
      year={2026},
      journal={arXiv:2605.11661},
      doi = {10.48550/arXiv.2605.11661}
}

@article{Si2015gap,
author = {Guo, Yaguang and Wang, Qian and Kawazoe, Yoshiyuki and Jena, Puru},
year = {2015},
month = {09},
pages = {14342},
title = {A New Silicon Phase with Direct Band Gap and Novel Optoelectronic Properties},
volume = {5},
journal = {Scientific Reports},
doi = {10.1038/srep14342}
}

@article{SiPhonon2012,
  title = {Phonon-Assisted Optical Absorption in Silicon from First Principles},
  author = {Noffsinger, Jesse and Kioupakis, Emmanouil and Van de Walle, Chris G. and Louie, Steven G. and Cohen, Marvin L.},
  journal = {Physical Review Letters},
  volume = {108},
  issue = {16},
  pages = {167402},
  numpages = {5},
  year = {2012},
  month = {04},
  publisher = {American Physical Society},
  doi = {10.1103/PhysRevLett.108.167402},
  url = {https://link.aps.org/doi/10.1103/PhysRevLett.108.167402}
}

@article{SiWavelength1970,
  title = {Wavelength-Modulation Spectra of Some Semiconductors},
  author = {Zucca, Ricardo R. L. and Shen, Y. R.},
  journal = {Physical Review B},
  volume = {1},
  issue = {6},
  pages = {2668--2676},
  numpages = {0},
  year = {1970},
  month = {03},
  publisher = {American Physical Society},
  doi = {10.1103/PhysRevB.1.2668},
  url = {https://link.aps.org/doi/10.1103/PhysRevB.1.2668}
}

@article{geuzaine2009gmsh,
author = {Geuzaine, Christophe and Remacle, Jean-François},
year = {2009},
month = {09},
pages = {1309 - 1331},
title = {Gmsh: A {3-D} Finite Element Mesh Generator with Built-in Pre- and Post-Processing Facilities},
volume = {79},
journal = {International Journal for Numerical Methods in Engineering},
doi = {10.1002/nme.2579}
}

@article{Polman2019_energy_momentum_CL,
    author = {Coenen, Toon and Polman, Albert},
    title = {Energy-Momentum Cathodoluminescence Imaging of Anisotropic
Directionality in Elliptical Aluminum Plasmonic Bullseye Antennas},
    journal = {ACS Photonics},
    volume = {6},
    number = {2},
    pages = {573-580},
    year = {2019},
    month = {12},
    issn = {2330-4022},
    doi = {10.1021/acsphotonics.8b01711},
    url = {https://doi.org/10.1021/acsphotonics.8b01711},
    eprint = {https://pubs.acs.org/apchd5/article-pdf/6/2/573/17573065/ph8b01711.pdf}
}

@article{Talebi_2013,
doi = {10.1088/1367-2630/15/5/053013},
url = {https://doi.org/10.1088/1367-2630/15/5/053013},
year = {2013},
month = {05},
publisher = {IOP Publishing},
volume = {15},
number = {5},
pages = {053013},
author = {Talebi, Nahid and Sigle, Wilfried and Vogelgesang, Ralf and van Aken, Peter},
title = {Numerical simulations of interference effects in photon-assisted electron energy-loss spectroscopy},
journal = {New Journal of Physics}
}

@article{Chahshouri_2022,
doi = {10.1088/1361-6463/ac453a},
url = {https://doi.org/10.1088/1361-6463/ac453a},
year = {2022},
month = {jan},
publisher = {IOP Publishing},
volume = {55},
number = {14},
pages = {145101},
author = {Chahshouri, Fatemeh and Taleb, Masoud and Diekmann, Florian K and Rossnagel, Kai and Talebi, Nahid},
title = {Interaction of excitons with Cherenkov radiation in WSe2 beyond the non-recoil approximation},
journal = {Journal of Physics D: Applied Physics}
}

@article{Yamamoto01091991,
author = {Naoki Yamamoto and Hiroki Sugiyama},
title = {Cherenkov and transition radiation generated in 200 kV electron microscope},
journal = {Radiation Effects and Defects in Solids},
volume = {117},
number = {1-3},
pages = {5--10},
year = {1991},
publisher = {Taylor \& Francis},
doi = {10.1080/10420159108220585},
URL = {https://doi.org/10.1080/10420159108220585},
eprint = {https://doi.org/10.1080/10420159108220585}
}

@article{Yamamoto1996_CR_TR_thin_plate,
author = {Yamamoto, Naoki and Sugiyama, Hiroki and Toda, Akio},
year = {1996},
month = {10},
pages = {2279-2301},
title = {Cherenkov and Transition Radiation from Thin Plate Crystals Detected in the Transmission Electron Microscope},
volume = {452},
journal = {Proceedings of the Royal Society of London A},
doi = {10.1098/rspa.1996.0122}
}

@article{Yamamoto1996_TR_imaging,
    author = {Yamamoto, Naoki and Toda, Akio and Axaya, Katsuhiro},
    title = {Imaging of Transition Radiation from Thin Films on a Silicon Substrate Using a Light Detection System Combined with TEM},
    journal = {Journal of Electron Microscopy},
    volume = {45},
    number = {1},
    pages = {64-72},
    year = {1996},
    month = {02},
    issn = {0022-0744},
    doi = {10.1093/oxfordjournals.jmicro.a023414},
    url = {https://doi.org/10.1093/oxfordjournals.jmicro.a023414},
    eprint = {https://academic.oup.com/jmicro/article-pdf/45/1/64/2649984/45-1-64.pdf}
}

@article{StoegerPollach,
author = {Stöger-Pollach, M. and Kachtík, Lukáš and Miesenberger, Bernhard and Retzl, Philipp},
year = {2017},
pages = {31-35},
title = {Transition Radiation in EELS and Cathodoluminescence},
volume = {173},
journal = {Ultramicroscopy},
doi = {10.1016/j.ultramic.2016.11.020}
}

@article{Haslinger2022,
title = {Discrimination of coherent and incoherent cathodoluminescence using temporal photon correlations},
journal = {Ultramicroscopy},
volume = {241},
pages = {113594},
year = {2022},
issn = {0304-3991},
doi = {https://doi.org/10.1016/j.ultramic.2022.113594},
url = {https://www.sciencedirect.com/science/article/pii/S0304399122001127},
author = {Michael Scheucher and Thomas Schachinger and Thomas Spielauer and Michael Stöger-Pollach and Philipp Haslinger}
}

@article{Abajo2004_Boundary_effects_CR,
  title = {Boundary effects in {Cherenkov} radiation},
  author = {Garc\'{\i}a de Abajo, F. J. and Rivacoba, A. and Zabala, N. and Yamamoto, N.},
  journal = {Physical Review B},
  volume = {69},
  issue = {15},
  pages = {155420},
  numpages = {12},
  year = {2004},
  month = {Apr},
  publisher = {American Physical Society},
  doi = {10.1103/PhysRevB.69.155420},
  url = {https://link.aps.org/doi/10.1103/PhysRevB.69.155420}
}

@article{Varkentina2022,
author = {Nadezda Varkentina  and Yves Auad  and Steffi Y. Woo  and Alberto Zobelli  and Laura Bocher  and Jean-Denis Blazit  and Xiaoyan Li  and Marcel Tencé  and Kenji Watanabe  and Takashi Taniguchi  and Odile Stéphan  and Mathieu Kociak  and Luiz H. G. Tizei },
title = {Cathodoluminescence excitation spectroscopy: Nanoscale imaging of excitation pathways},
journal = {Science Advances},
volume = {8},
number = {40},
pages = {eabq4947},
year = {2022},
doi = {10.1126/sciadv.abq4947},
URL = {https://www.science.org/doi/abs/10.1126/sciadv.abq4947},
eprint = {https://www.science.org/doi/pdf/10.1126/sciadv.abq4947}}

@article{Aharon2026,
author = {Aharon, Hadar and Barkay, Zahava and Meuret, Sophie and Kfir, Ofer},
title = {Cathodoluminescence Enhancement Mechanisms in Silica Microspheres},
journal = {Nanophotonics},
volume = {15},
number = {6},
pages = {e70055},
doi = {https://doi.org/10.1002/nap2.70055},
url = {https://onlinelibrary.wiley.com/doi/abs/10.1002/nap2.70055},
eprint = {https://onlinelibrary.wiley.com/doi/pdf/10.1002/nap2.70055},
year = {2026}
}

@article{Sannomiya2021,
    author = {Matsukata, Taeko and García de Abajo, F. Javier and Sannomiya, Takumi},
    title = {Chiral Light Emission from a Sphere Revealed by Nanoscale Relative-Phase Mapping},
    journal = {ACS Nano},
    volume = {15},
    number = {2},
    pages = {2219-2228},
    year = {2021},
    month = {},
    issn = {1936-0851},
    doi = {10.1021/acsnano.0c05624},
    url = {https://doi.org/10.1021/acsnano.0c05624},
    eprint = {https://pubs.acs.org/ancac3/article-pdf/15/2/2219/16501082/nn0c05624.pdf},
}

@article{Sannomiya2025,
    author = {Adachi, Yoshikazu and Machfuudzoh, Izzah and Kubota, Tetsuya and Yanagimoto, Sotatsu and Sugimoto, Hiroshi and Fujii, Minoru and Sannomiya, Takumi},
    title = {Temperature
Measurement in Transmission Electron Microscope
Using Cathodoluminescence of Whispering Gallery Mode},
    journal = {ACS Photonics},
    volume = {12},
    number = {8},
    pages = {4737-4744},
    year = {2025},
    month = {08},
    issn = {2330-4022},
    doi = {10.1021/acsphotonics.5c01418},
    url = {https://doi.org/10.1021/acsphotonics.5c01418},
    eprint = {https://pubs.acs.org/apchd5/article-pdf/12/8/4737/42317044/ph5c01418.pdf},
}

@article{Sannomiya2023,
    author = {Machfuudzoh, Izzah and Hinamoto, Tatsuki and García de Abajo, F. Javier and Sugimoto, Hiroshi and Fujii, Minoru and Sannomiya, Takumi},
    title = {Visualizing
the Nanoscopic Field Distribution of Whispering-Gallery
Modes in a Dielectric Sphere by Cathodoluminescence},
    journal = {ACS Photonics},
    volume = {10},
    number = {5},
    pages = {1434-1445},
    year = {2023},
    month = {03},
    issn = {2330-4022},
    doi = {10.1021/acsphotonics.3c00041},
    url = {https://doi.org/10.1021/acsphotonics.3c00041},
    eprint = {https://pubs.acs.org/apchd5/article-pdf/10/5/1434/3018604/ph3c00041.pdf},
}

@article{Sapienza2012,
  author       = {Sapienza, R. and Coenen, T. and Renger, J. and Kuttge, M. and van Hulst, N. F. and Polman, A.},
  title        = {Deep-subwavelength imaging of the modal dispersion of light},
  journal = {Nature Materials},
  year = {2012},
  month = {09},
  volume       = {11},
  number       = {9},
  pages        = {781-787},
  issn         = {1476-4660},
  doi          = {10.1038/nmat3402},
  url          = {https://doi.org/10.1038/nmat3402}
}

@article{soler_acsphot12,
author = {Soler, T. and Akerboom, E. and Stamatopoulou, P.~E. and Sugimoto, H. and Fujii, M. and Fiedler, S. and Polman, A.},
title = {Electron-energy dependent excitation and directional far-field radiation of resonant {Mie} modes in single {Si} nanospheres},
journal = {ACS Photonics},
volume = {32},
number = {8},
pages = {4161--4170},
year = {2025},
doi = {10.1021/acsphotonics.5c00173}
}

\end{document}


\begin{center}
{\Large\bfseries Supporting Information}
\vspace{1.2cm}
\end{center}

\title{Time- and frequency-domain study for electron beams penetrating dielectric nanospheres: fingerprints of Cherenkov and transition radiation}

\author{Wenhua~Zhao\,\orcidlink{0009-0004-5721-607X}}
\email{Corresponding author: wzhao@physik.hu-berlin.de}
\affiliation{Max-Born-Institut, 12489 Berlin, Germany}
\affiliation{Humboldt-Universit\"at zu Berlin, Institut f\"ur Physik, AG Theoretische Optik and Photonik, 12489 Berlin, Germany}
\author{Christos~Tserkezis\,\orcidlink{0000-0002-2075-9036}}
\affiliation{POLIMA---Center for Polariton-driven Light-Matter Interactions, University of Southern Denmark, 5230 Odense M, Denmark}
\affiliation{Danish Institute for Advanced Study, University of Southern Denmark, 5230 Odense M, Denmark}
\author{N.~Asger~Mortensen\,\orcidlink{0000-0001-7936-6264}}
\affiliation{POLIMA---Center for Polariton-driven Light-Matter Interactions, University of Southern Denmark, 5230 Odense M, Denmark}
\affiliation{Danish Institute for Advanced Study, University of Southern Denmark, 5230 Odense M, Denmark}
\author{Kurt~Busch\,\orcidlink{0000-0003-0076-8522}}
\email{Corresponding author: kurt.busch@physik.hu-berlin.de}
\affiliation{Humboldt-Universit\"at zu Berlin, Institut f\"ur Physik, AG Theoretische Optik and Photonik, 12489 Berlin, Germany}
\affiliation{Max-Born-Institut, 12489 Berlin, Germany}

\maketitle

\vspace{2.5cm}

\textbf{Supporting Information contains:}

\vspace{0.5cm}
11 pages

\vspace{0.5cm}
10 figures

\vspace{0.5cm}
3 tables

\thispagestyle{empty}
\newpage



\setcounter{page}{1}

    \textbf{In this document, we provide technical details for the manuscript entitled "Time- and frequency-domain study for electron beams penetrating dielectric nanospheres: fingerprints of Cherenkov and transition radiation". This includes the material models for silicon and computational insights via DGTD. Next, a study of Cherenkov radiation in a homogeneous medium and transition radiation at an interface is included, followed by an analytical decomposition of the EEL spectrum of a silicon sphere into different contributions. In addition, the effect of different substrate materials on the CL spectrum of a silicon sphere is shown. Finally, we show the time-domain movie for the swift electron beam traversing a dielectric sphere at high speed beyond the Cherenkov threshold.}

\tableofcontents

\section{Silicon material data and consistency study}

To start, we recapitulate the silicon (Si) lattice structure and its energy dispersion relation. Si crystallizes in a face-centered cubic (fcc) lattice, and its primitive cell in reciprocal space (the Brillouin zone) is a body-centered cubic (bcc) lattice.
In the band structure, there exist band gaps between the valence and conduction bands, which result from the interaction of valence electrons with the ion cores of the Si crystal~\cite{SiliconSpringer}.
The existence of band gaps between the valence band and the conduction band distinguishes Si from metallic materials~\cite{Si2015gap,SiPhonon2012, SiWavelength1970}.

To further understand the electromagnetic interaction between swift electrons and Si nanoparticles (NPs), we examine the Si material model.
The material data for Si at room temperature up to 30\,eV is taken from~\cite{Franta2017,RefractiveIndexInfo}. Note that it is important to consider high-energy material data to correctly recover the time-domain dynamics, where the usual data at lower energy range (up to 6\,eV~\cite{Green}) is not sufficient for our purpose. In order to 
employ the material data to perform calculations in DGTD, we need 
to fit the measured data~\cite{Franta2017} via a Drude--Lorentz model~\cite{Si_lorentz2017},
\begin{equation} \label{eq:Drude_lorentz}
    \varepsilon (\omega)=\varepsilon_{\infty}-\frac{\omega_{\mathrm p}^2}{\omega^2+{\mathrm i}\gamma_{\mathrm D}\omega}+\sum_j \frac{f_j\omega_{0,j}^2}{\omega_{0,j}^2-\omega^2-{\mathrm i}\gamma_{\mathrm L,j}\omega},
\end{equation}
where $\varepsilon_{\infty}$ denotes the background permittivity, $\omega_{\mathrm p}$ and $\gamma_{\mathrm D}$ denote the bulk plasmon frequency and damping frequency in Drude term, while $\omega_{0,j}$ and $\gamma_{\mathrm L,j}$ denote the resonance frequencies and damping constants in Lorentz terms. $f_j$ gives the strength for each Lorentz pole. The fitted parameters are as follows: the background permittivity is $\varepsilon_\infty = 0.9087$. The
Drude contribution is characterized by $\omega_\mathrm{p} = 0.00264$\,eV and $\gamma_D = 9.58 \times 10^{-21} \approx 0$. The Lorentz poles are listed in Table~\ref{tab:Si_parameters}.
\begin{table}[!htb]
\centering
\caption{\label{tab:Si_parameters}
Parameters for the numerical fitting of the measured Si permittivity $\varepsilon$ from~\cite{Franta2017,RefractiveIndexInfo} via 8 Lorentz poles according to Eq.~\eqref{eq:Drude_lorentz}.}
\begin{tabular}{cccc}
Lorentz poles & Strength $f$ &$\hbar\omega_0$ (eV) & $\hbar\gamma$ (eV) \\
\hline
Lorentz 1 & $2.953$ & $4.236$ & $0.364$ \\
Lorentz 2 & $1.549$ & $5.309$ & $1.048$ \\
Lorentz 3 & $1.030$ & $3.392$ & $0.145$ \\
Lorentz 4 & $1.144$ & $3.714$ & $0.254$ \\
Lorentz 5 & $0.915$ & $3.524$ & $0.188$ \\
Lorentz 6 & $1.000$ & $4.485$ & $0.473$ \\
Lorentz 7 & $0.965$ & $7.006$ & $2.443$ \\
Lorentz 8 & $1.441$ & $3.948$ & $0.321$
\end{tabular}
\end{table}

\begin{figure}[!htb]
\centering
\includegraphics[trim = 18mm 80mm 15mm 45mm, clip, width=1.0
    \textwidth]{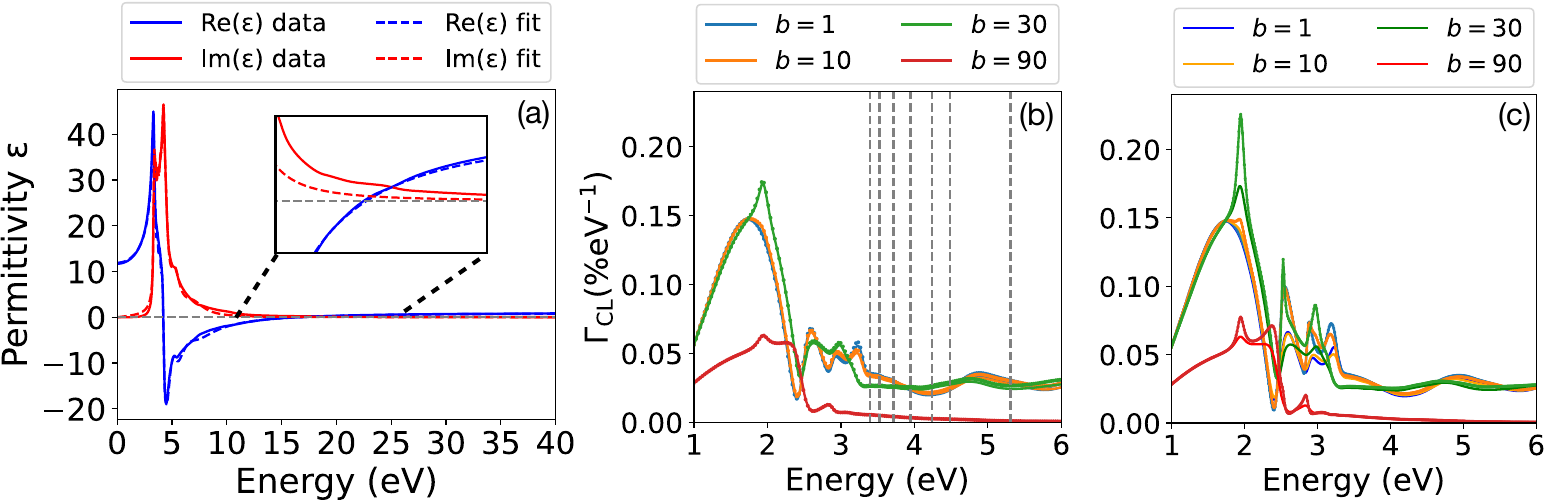}
    \caption{ \textbf{Silicon material data at room temperature and consistency study via DG and Mie.}
    \textbf{\pnl{a}} The comparison of the Si permittivity $\varepsilon$ from~\cite{Franta2017,RefractiveIndexInfo} as dashed lines and the fitting data using Drude--Lorentz model via 8 Lorentz poles according to Eq.~\eqref{eq:Drude_lorentz}. 
    \textbf{\pnl{b}} The comparison between DG simulations (solid curves) using fitted material data and analytical Mie-solutions~\cite{Elli_Mie_code} (dotted-lines) using fitted material data for different electron trajectories with $\beta=0.33$.
    \textbf{\pnl{c}} Same as \pnl{b}, but the Mie-solutions are calculated using experimental Si material data~\cite{Franta2017}.
    }
    \label{fig:Si_material_data}
\end{figure}

Specifically, both Drude term and Lorentz poles are essential to achieve an optimum fitting result. Note that by the fitting we constrained the parameters to be positive for a realistic physical material model. The Lorentz oscillators represent effective bound-electron and interband transitions in Si~\cite{Si2015gap,SiPhonon2012, SiWavelength1970}; although individual oscillators should not necessarily be interpreted as single physical resonances.
The Drude term at lower energy, with negligible loss, is a consequence of the fitting process used to capture a slowly varying background in the permittivity around this energy and does not correspond to a physical excitation in Si. In Fig.~\ref{fig:Si_material_data}\pnl{a}, we show the comparison of the Si permittivity $\varepsilon$ from~\cite{Franta2017,RefractiveIndexInfo} and the fitting data using Drude--Lorentz model via 8 Lorentz poles according to Eq.~\eqref{eq:Drude_lorentz}.

With the material fitting data at hand, we perform the DGTD simulation for a swift electron with speed $\beta=0.33$ penetrating a Si nano sphere of radius $R=79$\,nm at different impact parameters $b$.  
We begin by comparing the CL spectra obtained from DGTD simulations with the analytical Mie solutions~\cite{Elli_Mie_code} using fitted material data. The corresponding results are shown in Fig.~\ref{fig:Si_material_data}\pnl{b}. We note a very good match between the analytical and numerical results. To quantify the difference between the formally exact Mie and the computational DG results, we define the relative error as, 
\begin{equation} \label{eq:relative_error}
    \mathrm{Rel.\,error} = \frac{\int d\omega \left| \Gamma_\mathrm{Mie} (\omega) - \Gamma_\mathrm{DGTD}(\omega) \right|}{\int d\omega \left| \Gamma_\mathrm{Mie}(\omega)\right|} 
    ,
\end{equation}
with which we calculate the relative error between the analytic Mie result and the DGTD result. The relative errors are shown in Table~\ref{tab:relative_error_Si}, with the integrals along the energy window $0.1-6$\,eV. 

\begin{table}[!htb]
\centering
\caption{\label{tab:relative_error_Si}
Relative error according to Eq.~\eqref{eq:relative_error} between the analytic Mie calculations and the DGTD calculations of $\Gamma_\mathrm{CL}$ for a swift electron interacting with Si nanosphere for different trajectories at an electron velocity $\beta=0.33$. We consider the fitted Drude--Lorentz model here.}
\begin{tabular}{ccc}
 $b$ (nm) &$\Gamma_\mathrm{error}$ (\%) \\
\hline
1&  2.6   \\
10& 2.5   \\
30&  2.6  \\
90&  0.4
\end{tabular}
\end{table}

We only calculate the relative errors for the low-energy range because, for the mesh size $h_\mathrm{max}=5$\,nm and a fixed computation time of approximately 200\,fs in the DG scheme, we expect convergent behavior in the low-energy range. To achieve convergence of the spectrum in the high-energy range, a smaller mesh size and longer simulation time would be required, which is not necessary for our purposes. The small relative errors indicate a correct implementation of the material model in the DGTD framework.

To ensure that the fitted material model reflects the real Si material properties, we perform a comparison between DG simulations (solid curves) using fitted material data and analytical Mie solutions (dotted lines) using experimental Si material data~\cite{Franta2017}, as shown in Fig.~\ref{fig:Si_material_data}\pnl{c}. We observe differences between them, with the relative errors shown in Table~\ref{tab:relative_error_Si_DG_Mie_exp}. However, the main resonance peaks are well captured by the Drude--Lorentz fit.

\begin{table}[!htb]

\centering
\caption{\label{tab:relative_error_Si_DG_Mie_exp}
Relative error according to Eq.~\eqref{eq:relative_error} between the analytic and the DGTD calculations of $\Gamma_\mathrm{CL}$ for a swift electron interacting with Si nanosphere for different trajectories at an electron velocity $\beta=0.33$. For DG we consider the fitted Drude--Lorentz model, while for Mie we consider experimental Si material data~\cite{Franta2017}.}
\begin{tabular}{ccc}
 $b$ (nm) &$\Gamma_\mathrm{error}$ (\%) \\
\hline
1&  6.9   \\
10& 7.2   \\
30&  9.1  \\
90&  8.6
\end{tabular}
\end{table}

\begin{figure}[!htb]
\centering
\includegraphics[trim = 0mm 83mm 0mm 50mm, clip, width=1.0\textwidth]{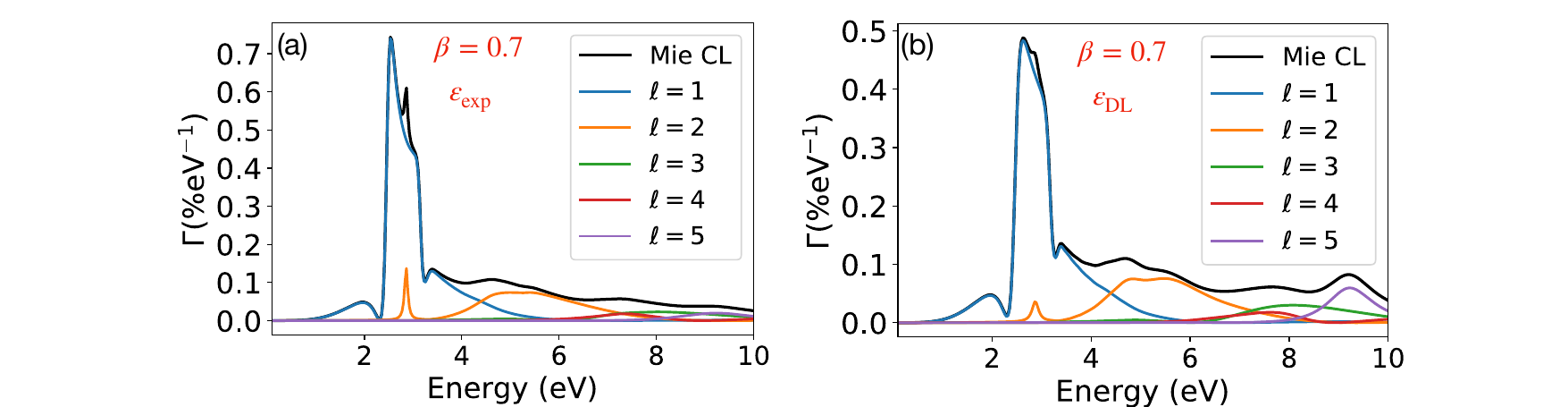}
    \caption{ \textbf{Mie calculations for CL of a Si sphere with different material data.}
    We consider a Si sphere of $R=79$\,nm penetrated by a swift electron beam with $\beta=0.7$ at impact distance $b=1$\,nm.
    \textbf{\pnl{a}} Mie decompositions of CL for Si with experimental permittivity $\varepsilon$~\cite{Franta2017,RefractiveIndexInfo}.
    \textbf{\pnl{b}} Mie decompositions of CL for Si with fitted data using Drude--Lorentz model via 8 Lorentz poles according to Eq.~\eqref{eq:Drude_lorentz}.
    }
    \label{fig:Si_mie_07c}
\end{figure}

Furthermore, for an electron velocity of $\beta = 0.7$, we present the Mie decompositions in Fig.~\ref{fig:Si_mie_07c}, using the experimental permittivity $\varepsilon$~\cite{Franta2017,RefractiveIndexInfo}, as shown in Fig.~\ref{fig:Si_mie_07c}\pnl{a}, and fitted data using the Drude--Lorentz model with eight Lorentz poles according to Eq.~\eqref{eq:Drude_lorentz}, as shown in Fig.~\ref{fig:Si_mie_07c}\pnl{b}.
We study the case where a Si sphere with radius $R=79$\,nm is penetrated by a swift electron beam with $\beta = 0.7$ at an impact parameter $b = 1$\,nm. Specifically, at low energies around 2.5--3\,eV, we observe a clear peak splitting for both both material data/modeling.

%

\section{Cherenkov radiation}

\begin{figure}[!htb]
\centering
\includegraphics[trim = 0mm 65mm 0mm 58mm, clip, width=1.0
    \textwidth]{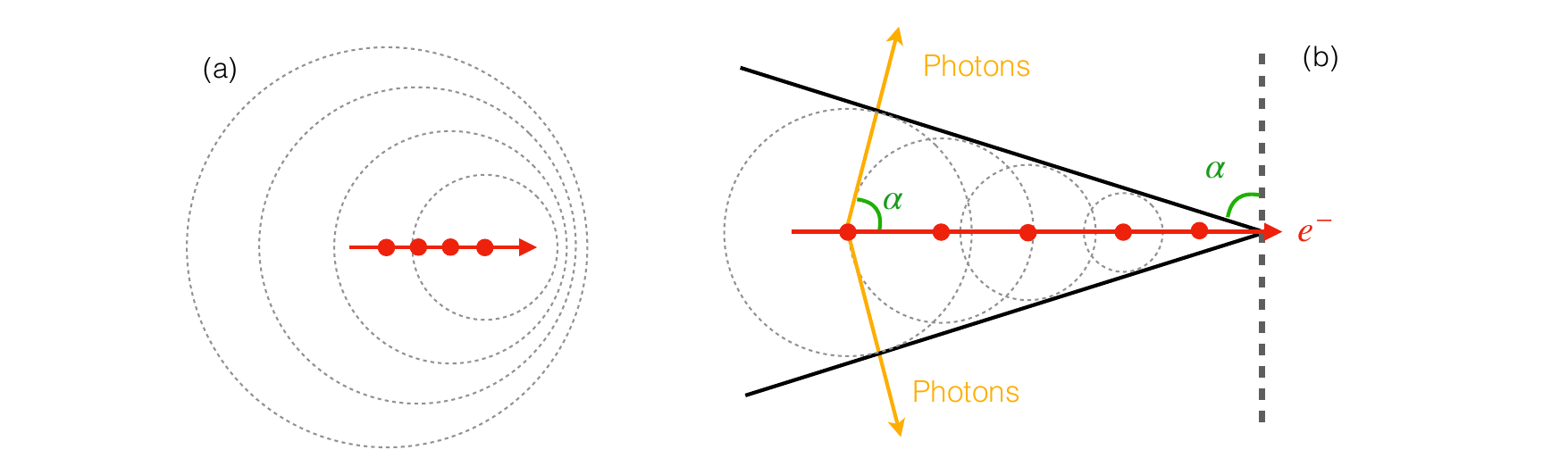}
    \caption{\textbf{Illustration of the Cherenkov radiation (CR). The velocity of the charged particle, denoted by red points, is given by $v_e=\beta c$. The emission angle $\alpha$ is given by Eq.~\eqref{eq:CR_angle}}.
    \textbf{\pnl{a}} The CR condition is not fulfilled, where $v_e < c/n$.
    \textbf{\pnl{b}}~The CR condition is fulfilled, where $v_e > c/n$. We observe the spherical waves cross with each other and build the "shock front", denoted with black curves.
    }
    \label{fig:Si_CR_Illustration}
\end{figure}

The Vavilov--Cherenkov effect~\cite{TR_Ginzburg1996} describes the fact that electromagnetic radiation (Cherenkov radiation) could be emitted by a charged particle moving uniformly through a medium if the Cherenkov condition
\begin{equation} \label{eq:CR_condition}
    v_e > \frac{c}{n(\omega)}
\end{equation}
is satisfied. Here, $v_e=\beta c$ is the velocity of the charged particle. This means that when the velocity of the charged particle exceeds the phase velocity of light in the medium, Cherenkov radiation (CR) is emitted with a continuous spectrum and specific angular distribution. To characterize the angular distribution, we consider the energy-momentum conservation before and after the emission of CR
\begin{align}\label{eq:CR_E_P_conservation}
    E_0 &= E_1 + \hbar \omega \\ \nonumber
    \mathbf{P}_0 &= \mathbf{P}_1 + \hbar \mathbf{k},
\end{align} 
with the index $0$ referring to the energy and momentum before the emission of CR, and the index $1$ after that. $\hbar \omega$ gives the energy of emitted photons. Using the relativistic energy-momentum relation for the charged particle, $E^2=m^2c^4+c^2p^2$, we obtain
\begin{align} \label{eq:CR_E_P_conservation2}
    &p_1^2 = p_0^2 + p_\mathrm{photon}^2 - 2p_0p_\mathrm{photon} \, \cos\alpha \\ \nonumber
    &\sqrt{p_0^2 c^2 + m^2c^4} = \sqrt{p_1^2c^2 + m^2c^4} + \hbar \omega
\end{align}
By combining the two expressions in Eq.~\eqref{eq:CR_E_P_conservation2} and neglecting the small energy of the photons $\hbar \omega$ we obtain the spectral condition of CR,
\begin{equation} \label{eq:CR_angle}
    \cos\alpha \approx \frac{1}{\beta n},
\end{equation}
where the angle $\alpha$ gives the radiation direction of the emitted photons compared to the propagation direction of the charged particle, as shown in Fig.~\ref{fig:Si_CR_Illustration}. In our case, we consider a swift electron with relativistic velocities.

The radiated energy per unit time (power)~\cite{TR_Ginzburg1996} for a point charge $-e$ moving with velocity $v_e$ in a medium with a refraction index $n(\omega)$ is given by
\begin{equation} \label{eq:CR_differential_radiated_energy}
\frac{d\mathrm{E}^\mathrm{CR}}{dt}=\frac{e^{\,2} v_e}{4 \pi \varepsilon_0 c^{2} }
\int_{\,c/[n(\omega)v_e]\le 1}
\left(1 - \frac{c^{2}}{n^{2}(\omega)v_e^{2}}
\right) \omega\, d\omega,
\end{equation}
where the integration is performed over $\omega$ where the Cherenkov condition in Eq.~\eqref{eq:CR_condition} is satisfied. Similarly to the definition of the EEL probability, we define here a Cherenkov radiation probability
\begin{equation} \label{eq:CR_probability}
\Gamma_\mathrm{CR}=\frac{e^{\,2}}{4 \pi \varepsilon_0 c^{2} \hbar^2}
\int_{\,c/[n(\omega)v_e]\le 1}
\left(1 - \frac{1}{n^{2}(\omega)\beta^{2}}
\right) \, dz,
\end{equation}
integrated over the entire duration of the interaction within the nanoparticle for the energy range where the Cherenkov condition Eq.~\eqref{eq:CR_condition} is satisfied. 

\begin{figure}[!htb]
\centering
\includegraphics[trim = 20mm 53mm 37mm 5mm, clip, width=1.0\textwidth]{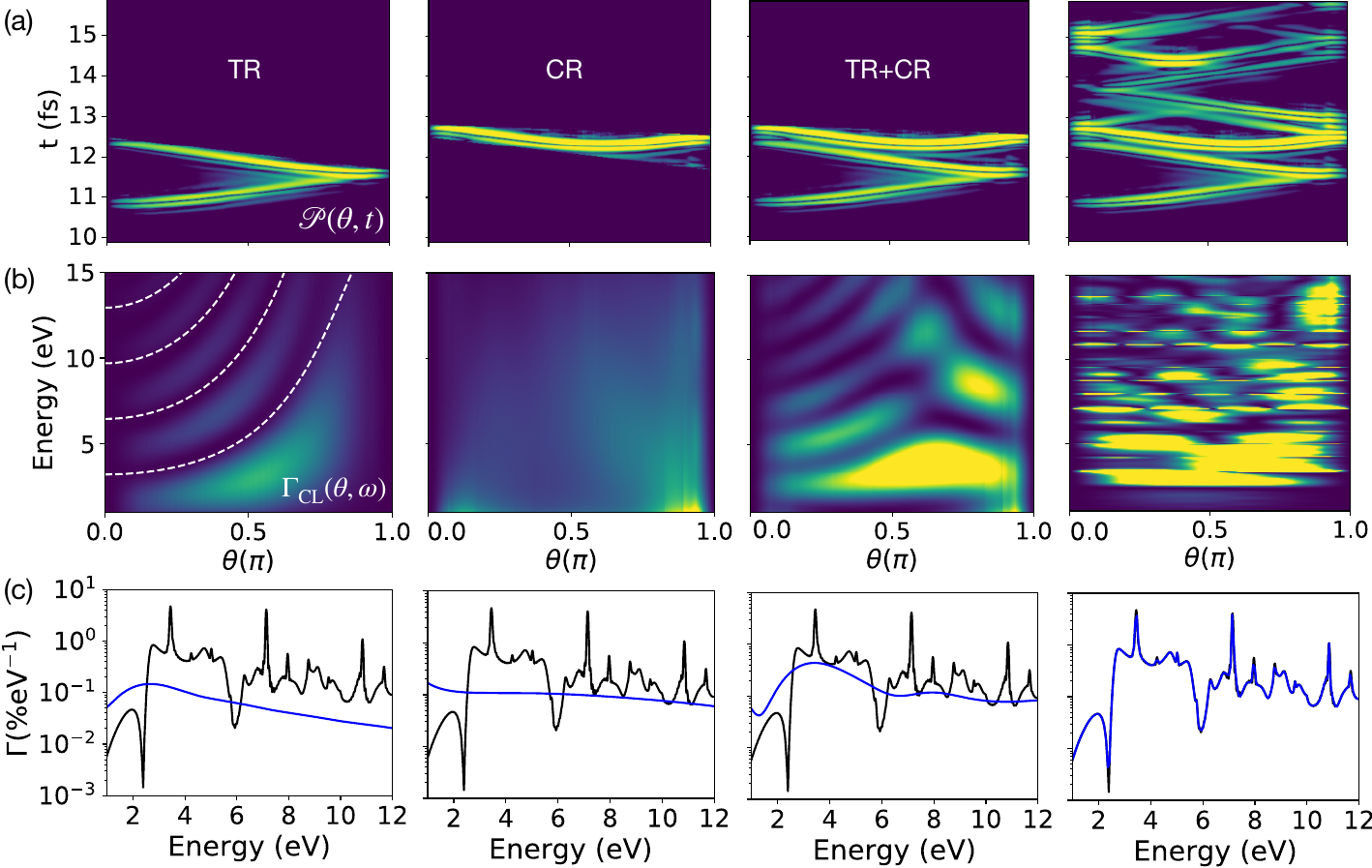}
    \caption{ \textbf{Decomposing TR and CR contributions in time and frequency via DGTD.}
    We consider a dielectric sphere of $R=79$\,nm with a constant permittivity $\varepsilon=16$ penetrated by a swift electron beam with $\beta=0.7$ at impact distance $b=0$\,nm.
    \textbf{\pnl{a}} We apply a mask in terms of time $t$ and angle $\theta$ to isolate the main TR/CR/TR+CR contributions in angle-resolved Poynting flux, respectively. 
   \textbf{\pnl{b}} We take the fields $\mathbf{E}$ and $\mathbf{H}$ corresponding to the mask in \pnl{a} and calculate the angle-resolved CL. The interference condition in Eq.~(8) of the main text for $m=1,\ldots,4$ is superimposed onto the map for TR as white-dashed curves.
   \textbf{\pnl{c}} By integrating \pnl{b} over angle $\theta$ we obtain the CL for each contribution, denoted by the blue curve. The black curve gives the total CL.
    }
    \label{fig:SI_CL_mask}
\end{figure}

As discussed in the main text, we can perform decomposition of TR and CR contributions---here for a dielectric sphere of $R=79$\,nm with a constant permittivity $\varepsilon=16$ penetrated by a swift electron beam with $\beta=0.7$ at impact distance $b=0$---as shown in Fig.~\ref{fig:SI_CL_mask}. In \pnl{a}, we apply a mask in terms of time $t$ and angle $\theta$ to isolate the main TR/CR/TR+CR contributions in angle-resolved Poynting flux, respectively. The last column corresponds to the full signal where we include all contributions in time. Next, we take the fields $\mathbf{E}$ and $\mathbf{H}$ corresponding to the mask in \pnl{a} and calculate the angle-resolved CL, as displayed in row \pnl{b}. Specifically, we obtain the characteristic fringes when only considering TR, as indicated by the double-slit interference pattern according to Eq.~(8) of the main text, while the CR gives a smooth background. When including both TR and CR, they interplay with each other and build interference features. Finally, when including all contributions, the complex CL map appears. By integrating \pnl{b} over angle $\theta$ we obtain the CL spectrum for each contribution, denoted by the blue curves in \pnl{c}. Interestingly, the TR+CR contribution creates an envelope of the total CL spectrum, thereby shaping the main feature of the far-field radiation.

%

\section{Analytical EELS decomposition}

Apart from the emission of radiation, part of the energy transferred to the optical modes of the nanoparticle dissipates non-radiatively, owing to intrinsic material losses. The total energy loss can be calculated as the work done by the electron against the induced field along its entire trajectory. In a semiclassical description, the resulting electron energy-loss (EEL) probability~\cite{Abajo2021,Agwire2026} for the electron losing an energy $\hbar\omega$ is given by
\begin{equation} \label{EELS_general}
    \Gamma_{\mathrm{EEL}} (\omega) = 
    \frac{e}{\pi\hbar\omega} 
    \int dt \, 
    \mathrm{Re} \big\{ \mathrm{e}^{-i\omega t} \,
    \mathbf{v} \cdot 
    \mathbf{E}_{\mathrm{ind}} (\mathbf{r}_{e}, \omega) \big\}.
\end{equation}
The integral in Eq.~\eqref{EELS_general} can be decomposed into three terms~\cite{Elli_Wenhua2023,Elli_Mie_code}
\begin{equation} \label{EELS_decomposition}
    \Gamma_{\mathrm{EEL}} (\omega) = 
    \Gamma_{\mathrm{bulk}} (\omega) + 
    \Gamma_{\mathrm{surf}} (\omega) +  
    \Gamma_{\mathrm{Begr}} (\omega).
\end{equation}
Here, $\Gamma_{\mathrm{bulk}}$ is related to the bulk modes of the unbound medium, reduced by the Begrenzung term $\Gamma_{\mathrm{Begr}}$ that accounts for the presence of a boundary. The $\Gamma_{\mathrm{surf}}$ term contains the contribution from modes excited by the part of the electron trajectory lying externally to the NP, and is, thus, associated with surface excitations. Although the analytical decompositions do not necessarily correspond to the exact physical decompositions into bulk and surface parts, the formulations are helpful to estimate the weight of different contributions. 

\begin{figure}[!htb]
\centering
\includegraphics[trim = 5mm 97mm 0mm 30mm, clip, width=1.0
    \textwidth]{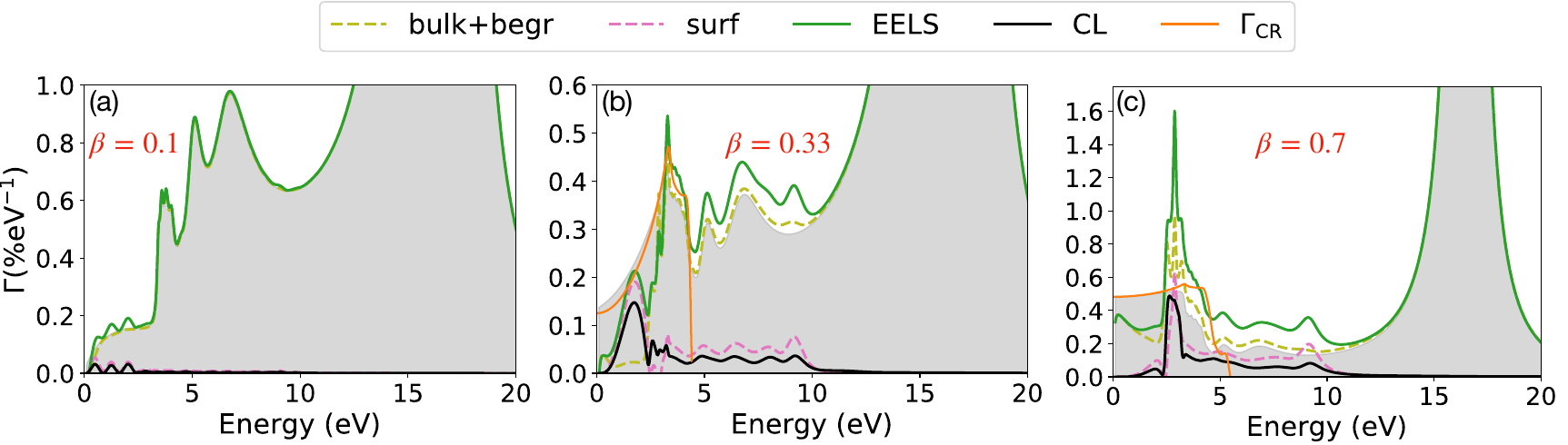}
    \caption{\textbf{Analytical EELS decompositions and study of the Cherenkov Radiation probability $\Gamma_\mathrm{CR}$ for different electron velocities.} We use the fitted measured Si permittivity $\varepsilon$ in Table~\ref{tab:Si_parameters} according to Eq.~\eqref{eq:Drude_lorentz}.
    We compare the EEL, CL and CR probability, where CR probability (orange) is calculated according to Eq.~\eqref{eq:CR_probability} by neglecting the imaginary part of the refractive index first. The gray-filled color corresponds to the analytical bulk contribution.
    \textbf{\pnl{a}} For electron velocity $\beta=0.1$.
    \textbf{\pnl{b--c}} Similar as \pnl{a}, but for electron velocity $\beta=0.33$ and $\beta=0.7$, respectively.}
    \label{fig:Si_CR_EELS}
\end{figure}

\section{Transition radiation for dielectric spheres modeled with constant permittivity}

 \begin{figure}[!htb]
\centering
\includegraphics[trim = 4mm 90mm 10mm 30mm, clip, width=1.0
    \textwidth]{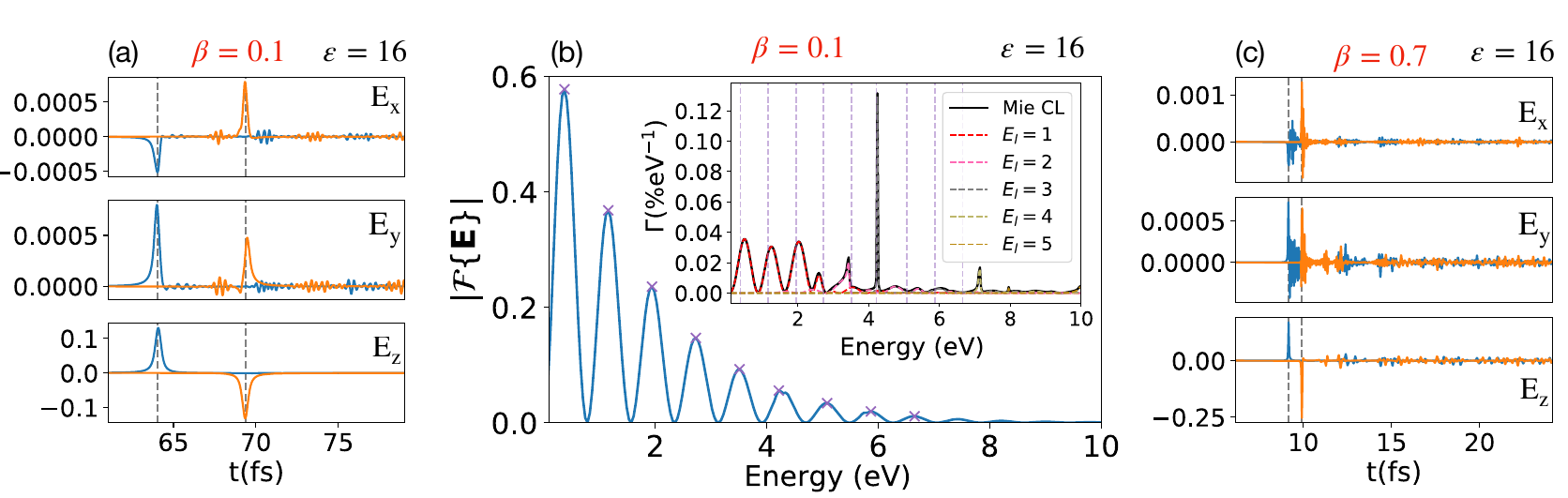}
    \caption{\textbf{TR for fast electron beams penetrating a dielectric sphere at different electron speeds.} 
   We consider a free-standing dielectric sphere modeled as $\varepsilon=16$ of radius $R=79$\,nm excited by a swift electron traveling downwards for penetrating trajectory $b=0$. 
   \textbf{\pnl{a}} We consider electron velocity $\beta=0.1$ and plot the near-field components $E_x(t)$, $E_y(t)$ and $E_z(t)$ at transition points. 
   The gray dashed-lines denote the arriving time of the electron at the two transition points.
   \textbf{\pnl{b}} The fields at two transition points in \pnl{a} are added together and Fourier transformed into frequency domain $|\mathcal{F} \{\mathbf{E}\}|$. The inset shows an analytical Mie decomposition.
   \textbf{\pnl{c}} Similar as \pnl{a} but for $\beta=0.7$.
   }
    \label{fig:Si_time_E_FT_eps16}
\end{figure}

We study TR excitation in a dielectric sphere with $\varepsilon = 16$, excited by a swift electron beam with a penetrating trajectory at $b = 0$. The radius of the sphere is $R = 79$\,nm. For lower electron velocities below the Cherenkov threshold, TR dominates the excitation channels; therefore, we choose $\beta = 0.1$ to study TR in this model.
Fig.~\ref{fig:Si_time_E_FT_eps16}\pnl{a} shows the three components of the electric field $\mathbf{E}$ at the transition points defined in Fig.~5 of the main text. We notice that the $E_z$ component is much stronger than the other components, and the delta-like excitation at the transition time points indicates strong TR excitation. To obtain the interference pattern, we coherently add the fields at the two transition points and Fourier transform them into the frequency domain, as shown in Fig.~\ref{fig:Si_time_E_FT_eps16}\pnl{b}. The interference peaks are denoted by purple crosses. The inset shows an analytical multipole decomposition, with the peaks identified above superimposed as purple vertical dashed lines. We observe very good agreement between them.
For comparison, we also include the field dynamics for a higher electron velocity of $\beta = 0.7$, above the Cherenkov threshold, as shown in Fig.~\ref{fig:Si_time_E_FT_eps16}\pnl{c}. There, we observe a much shorter pulse width than in the case with $\beta = 0.1$ shown in Fig.~\ref{fig:Si_time_E_FT_eps16}\pnl{a}. This is related to the incident fields carried by a relativistic electron, which exhibit shorter pulses at higher electron velocities~\cite{Maciel_Efield_electron2019,paper149,Jackson1998}.

\section{Silicon material model with less Lorentz-poles}

\begin{figure}[!htb]
\centering
\includegraphics[trim = 0mm 85mm 0mm 50mm, clip, width=1.0
    \textwidth]{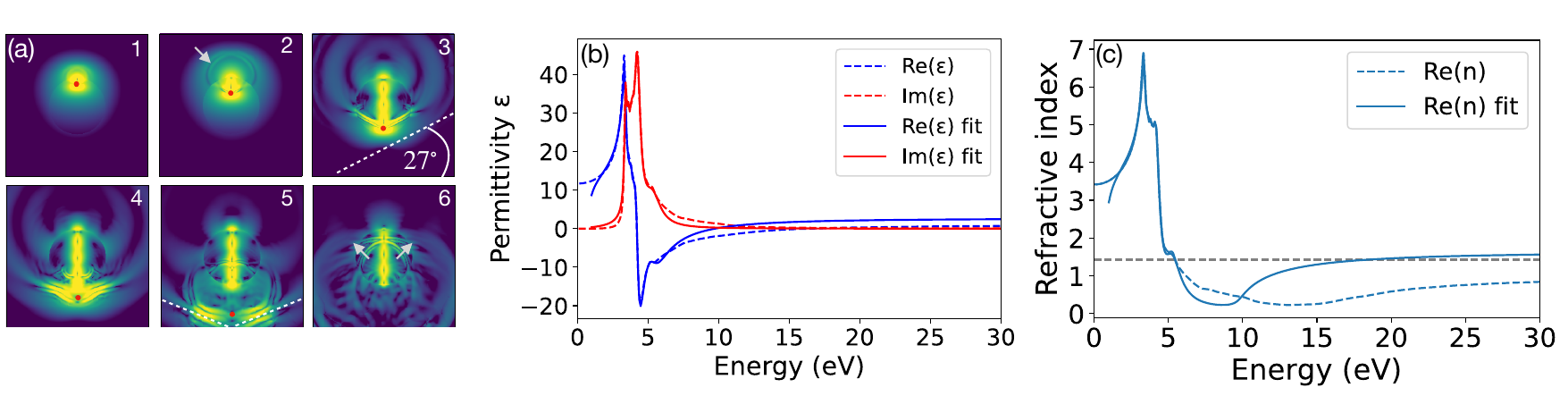}
    \caption{\textbf{Numerical simulation of the interaction between a fast electron beam and a Si sphere modeled via 5 Lorentz-poles.} 
   We consider a free-standing Si sphere modeled via 5 Lorentz-poles of radius $R=79$\,nm excited by a swift electron traveling downwards with velocity $\beta=0.7$ for penetrating trajectory $b=0$. 
   \textbf{\pnl{a}} Dynamics of the induced fields $|\mathbf{E}|$ at $xz$-cutplane at different time steps. The white-dashed lines correspond to a fitting of the CR front observed in step \pnl{a-3} and \pnl{a-5}.
   \textbf{\pnl{b}} Compare Si permittivity from~\cite{Franta2017} and fitted permittivity via 5 Lorentz-poles.
   \textbf{\pnl{c}} Similar as \pnl{b}, but we compare the real part of the refractive index $n$. The gray dashed line indicates the Cherenkov threshold for electron velocity $\beta=0.7$.}
    \label{fig:Si_time_E_FT_5Lpole}
\end{figure}

The time-domain simulation using DG is strongly influenced by the material modeling. To obtain reliable time-domain dynamics, a material model valid up to sufficiently high energies is required. In the main text, we presented a material model based on eight Lorentz poles, which accurately reproduces the material properties up to 30\,eV. Here, we demonstrate what may go wrong if Si is instead modeled using fewer Lorentz poles, as shown in Fig.~\ref{fig:Si_time_E_FT_5Lpole}.

In Fig.~\ref{fig:Si_time_E_FT_5Lpole}\pnl{a} we show the dynamics of the induced fields $|\mathbf{E}|$ at the $xz$-cutplane. 
In Fig.~\ref{fig:Si_time_E_FT_5Lpole}\pnl{a-3}, we observe the shock front resulting from CR, with a front angle of $27^\circ$, which is much smaller than in the case with $\varepsilon = 16$. The reason is that, in real Si, the additional Lorentz poles introduce significant damping into the material model, which slows down the waves at the relevant energies, such that these waves do not contribute to the shock front.
As a result, only the surviving waves contribute to the overall front, leading to a smaller front angle. In Fig.~\ref{fig:Si_time_E_FT_5Lpole}\pnl{a-4}, we observe that as the front reaches the spherical boundary, part of it survives total internal reflection and is transmitted into vacuum.
In Fig.~\ref{fig:Si_time_E_FT_5Lpole}\pnl{a-5}, we observe the escaped wave packets propagating into the far field while simultaneously spreading out. Meanwhile, in Fig.~\ref{fig:Si_time_E_FT_5Lpole}\pnl{a-6}, we observe that part of the front is reflected (gray arrows) and propagates in the bulk, where it is subsequently reflected and transmitted at the opposite boundary. The tightly spaced out-coupled fronts indicate a high-energy (around 20\,eV) out-coupling of the Cherenkov front into the far field.

This observation demonstrates a much stronger Cherenkov effect than Fig.~6 in the main text, because the material model used here, while correctly reproducing the low-energy data, deviates from the experimental data at higher energies. Specifically, the bulk plasmon energy between 16--17\,eV, defined as the crossing point between the real part of $\varepsilon$ and the $x$-axis, does not match the experimental value. The fitted $\mathrm{Re}[\varepsilon]$ is significantly higher than the experimental material data. This artifact affects the supported Cherenkov radiation. Fig.~\ref{fig:Si_time_E_FT_5Lpole}\pnl{c} shows that in the higher-energy range ($>20$\,eV), the Cherenkov condition is fulfilled for the fitted refractive index $n$, which leads to the observable Cherenkov front in Fig.~\ref{fig:Si_time_E_FT_5Lpole}\pnl{a}. In reality, the Cherenkov radiation should be strongly attenuated at high energies, as shown in Fig.~6 of the main text.

\section{TR at vacuum-dielectric interface when penetrated by fast electrons}

Here, we follow the Ginzburg and Frank theory of charged-particle transition radiation~\cite{TR_Ginzburg1996,TR_Chen2023}. For simplicity, we consider a vacuum–dielectric interface ($z = 0$). The moving electron current in the frequency domain is given by

\begin{equation}
\mathbf{J}(\mathbf r,\omega)
=- e\,\mathbf v\, \delta(\mathbf r_\perp)\, \mathrm{e}^{ \mathrm{i} \omega z/v_e }.
\end{equation}

The Helmholtz wave equation for the electromagnetic problem under study is given by
\begin{equation}
\nabla\times\nabla\times \mathbf E - \varepsilon(\mathbf r,\omega)\frac{\omega^2}{c^2}\mathbf E = \mathrm{i}\omega\mu_0 \mathbf J .
\end{equation}

We use a transverse Fourier representation for the electric fields
\begin{equation} \label{eq:TR_freq}
\mathbf E(\mathbf r_\perp,z,\omega)=\int \frac{d^2\mathbf k_\perp}{(2\pi)^2} \,\mathbf E(\mathbf k_\perp,z,\omega)\, \mathrm{e}^{\mathrm{i}\mathbf k_\perp\cdot\mathbf r_\perp}.
\end{equation}

The longitudinal wave vector is determined by
\begin{equation}
k_{z,j}=\sqrt{\varepsilon_{j}\frac{\omega^2}{c^2}-k_\perp^2},
\end{equation}
with $j$ indicating different regions and $\varepsilon_{j}$ the permittivity at different regions. We then make following radiation-field Ansatz for region~1 ($z<0$) and region~2 ($z>0$),
\begin{equation}
E^\mathrm{ind}_{z,1}(\mathbf k_\perp,z,\omega) =\frac{\mathrm{i} e}{\omega\varepsilon_0(2\pi)^3} a^{-}_{1,2} (\mathbf k_\perp,\omega)\, \mathrm{e}^{-\mathrm{i}k_{z,1}z}
\end{equation}
\begin{equation}
E^\mathrm{ind}_{z,2}(\mathbf k_\perp,z,\omega)
=\frac{\mathrm{i} e}{\omega\varepsilon_0(2\pi)^3} a^{+}_{1,2}(\mathbf k_\perp,\omega)\,
\mathrm{e}^{\mathrm{i}k_{z,2}z}.
\end{equation}

By enforcing the boundary conditions of the electromagnetic fields at the interface, we obtain the forward radiation coefficients $a^{+}_{1,2}$:
\begin{equation}
a^{+}_{1,2}
=\frac{\beta\,\frac{k_\perp^2 c^2}{\omega^2\varepsilon_{2}}
(\varepsilon_{1}-\varepsilon_{2})D_1^{(+)}
}{D_2^{(+)}D_3^{(+)}D_4}.
\end{equation}
And the backward radiation coefficient $a^{-}_{1,2}$:
\begin{equation}
a^{-}_{1,2}
=\frac{\beta\,\frac{k_\perp^2 c^2}{\omega^2\varepsilon_{1}}
(\varepsilon_{1}-\varepsilon_{2})D_1^{(-)}}{D_2^{(-)}D_3^{(-)}D_4}.
\end{equation}
\begin{align}
D_1^{(+)} &= 1-\beta^2\varepsilon_{2}
-\beta\frac{k_{z,1}}{\omega/c},
&\qquad
D_1^{(-)} &= 1-\beta^2\varepsilon_{1}
+\beta\frac{k_{z,2}}{\omega/c}, \\[4pt]
D_2^{(+)} &= 1-\beta^2\varepsilon_{2}
+\frac{k_\perp^2 v^2}{\omega^2},
&
D_2^{(-)} &= 1-\beta^2\varepsilon_{1}
+\frac{k_\perp^2 v^2}{\omega^2}, \\[4pt]
D_3^{(+)} &= 1-\beta\frac{k_{z,1}}{\omega/c},
&
D_3^{(-)} &= 1+\beta\frac{k_{z,2}}{\omega/c}.
\end{align}
and the common denominator $D_4=\varepsilon_{1}\frac{k_{z,2}}{\omega/c}+\varepsilon_{2}\frac{k_{z,1}}{\omega/c}$.
%
%
In the end, we reconstruct the time-domain formulation as,
\begin{equation} \label{eq:TR_time}
E_z^\mathrm{ind}(\mathbf r_\perp,z,t)=\int_{-\infty}^\infty d\omega\,
E_z^\mathrm{ind}(\mathbf r_\perp,z,\omega)\,\mathrm{e}^{-\mathrm{i}\omega t}.
\end{equation}

 \begin{figure}[!htb]
\centering
\includegraphics[trim = 20mm 70mm 40mm 60mm, clip, width=1.0
    \textwidth]{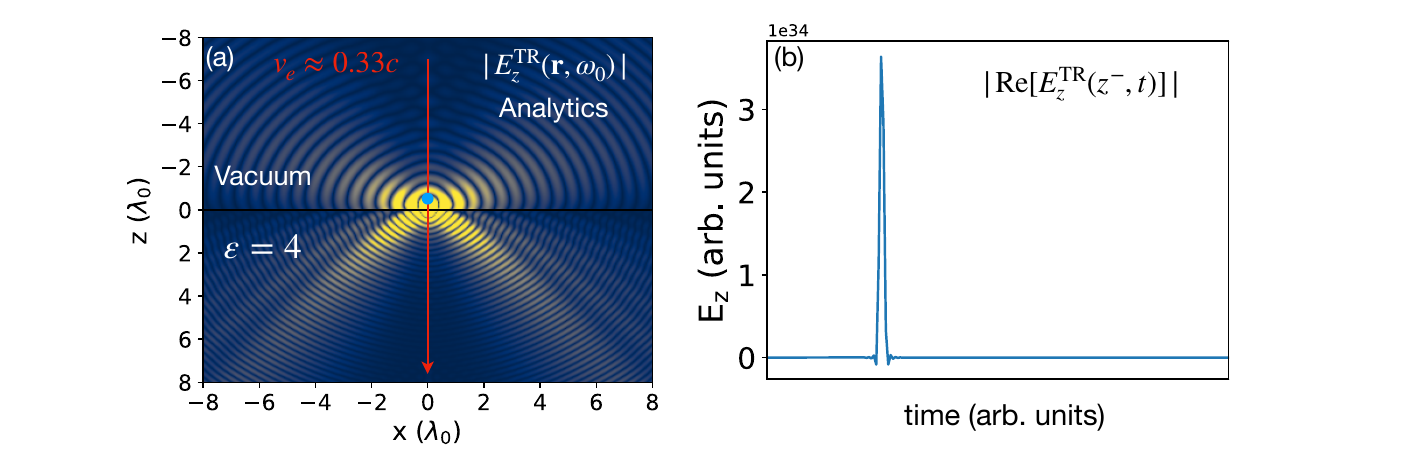}
    \caption{\textbf{Analytical TR radiation pattern in frequency and time domain.} We consider a vacuum-dielectric ($\varepsilon=4$) interface penetrated by a swift electron traveling with velocity $\beta=0.33$.
    \textbf{\pnl{a}} $E_z^\mathrm{ind}(x,z,\omega_0)$ following Eq.~\eqref{eq:TR_freq} for a specific frequency $\omega_0$ corresponding to wavelength $\lambda_0$.
    \textbf{\pnl{b}} $E_z^\mathrm{ind}(x=0,z=z^-,t)$ following Eq.~\eqref{eq:TR_time} for a position slightly before the transition point, marked as blue point in \pnl{a}.
    }
    \label{fig:Si_TR_slab_analytics}
\end{figure}

We now analyze the field patterns/dynamics of $E_z^\mathrm{ind}(x,z,\omega_0)$ at a specific frequency in the $xz$-cutplane according to Eq.~\eqref{eq:TR_freq}, as well as the time-dependent field $E_z^\mathrm{ind}(x=0,z=z^-,t)$ at a specific point $x=0,\, z=z^-$ according to Eq.~\eqref{eq:TR_time}, as shown in Fig.~\ref{fig:Si_TR_slab_analytics}.
We consider a vacuum-dielectric interface ($\varepsilon = 4$) penetrated by a swift electron traveling with velocity $\beta = 0.33$. We observe a similar impulse-like behavior of $E_z^\mathrm{ind}(x=0,z=z^-,t)$ as in Figs.~4 and 5 of the main text. Therefore, the delta-like impulse fields at the transition points in the main text are related to TR, whereas the additional oscillations in Fig.~5 of the main text are associated with damping introduced by the Lorentz poles in the Si material model.

\section{CL for silicon: effects of substrates}

\begin{figure}[!htb]
\centering
\includegraphics[trim = 0mm 55mm 0mm 50mm, clip, width=1.0
    \textwidth]{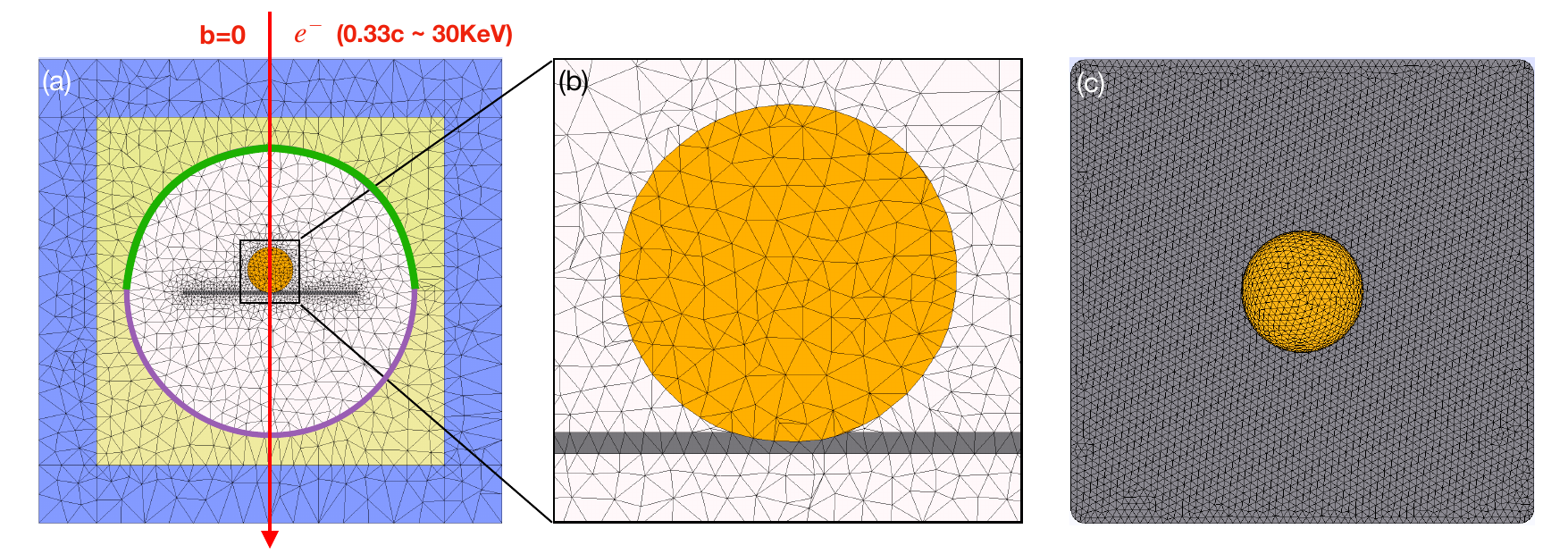}
    \caption{ \textbf{Mesh generated by GMSH~\cite{geuzaine2009gmsh} for a sphere deposited on a substrate.}
    \textbf{\pnl{a}} Mesh setup for the computation of CL spectra where a fast electron with velocity $\beta=0.33$ penetrates a Si sphere of radius $R=79$\,nm, with an impact parameter $b=0$, indicated as red arrows. The Si sphere is deposited on a substrate of width 0.6\,$\mathrm{\upmu m}$ and depth of $10$\,nm. The TF/SF surface is separated into two parts, the upper half sphere denoted with thick green line and the lower half sphere denoted with thick purple line.
    \textbf{\pnl{b}} The zoom of the contact region between Si sphere and the substrate. The sphere drops into the substrate and the penetration depth is $5$\,nm.
    \textbf{\pnl{c}} The mesh of the substrate with rounded edges.
    }
    \label{fig:Si_mesh_subs}
\end{figure}

In typical CL experiments, the Si sphere is deposited on a substrate; hence, we study the effects of different substrate materials on the acquired CL spectra of the Si sphere. These include Si (i.e., the same material as the sphere), silicon nitride ($\mathrm{Si_3N_4}$) with a constant permittivity $\varepsilon = 4$, and silver (Ag), described by a local Drude permittivity $\varepsilon(\omega) = 1 - \omega_\mathrm{p}^2/[\omega(\omega + \mathrm{i}\gamma)]$, with bulk plasmon energy $\hbar\omega_\mathrm{p} = 9.17$\,eV and damping constant $\hbar\gamma = 21$\,meV.
The mesh setup is shown in Fig.~\ref{fig:Si_mesh_subs}\pnl{a}, where we separate the TF/SF contour into an upper half-sphere and a lower half-sphere to mimic the experimental configuration. In Fig.~\ref{fig:Si_mesh_subs}\pnl{b}, we zoom in on the region where the sphere touches the substrate, where we embed the sphere 5\,nm into a 10\,nm deep substrate. This is necessary to avoid mesh errors at the contact point. In Fig.~\ref{fig:Si_mesh_subs}\pnl{c}, we show the substrate (gray region); for the same reason, we have rounded the edges to avoid mesh errors or excessively small elements, which could significantly reduce the allowable time step in DGTD and thus lead to unnecessarily long total simulation times.

\begin{figure}[!htb]
\centering
\includegraphics[trim = 16mm 67mm 32mm 72mm, clip, width=1.0
    \textwidth]{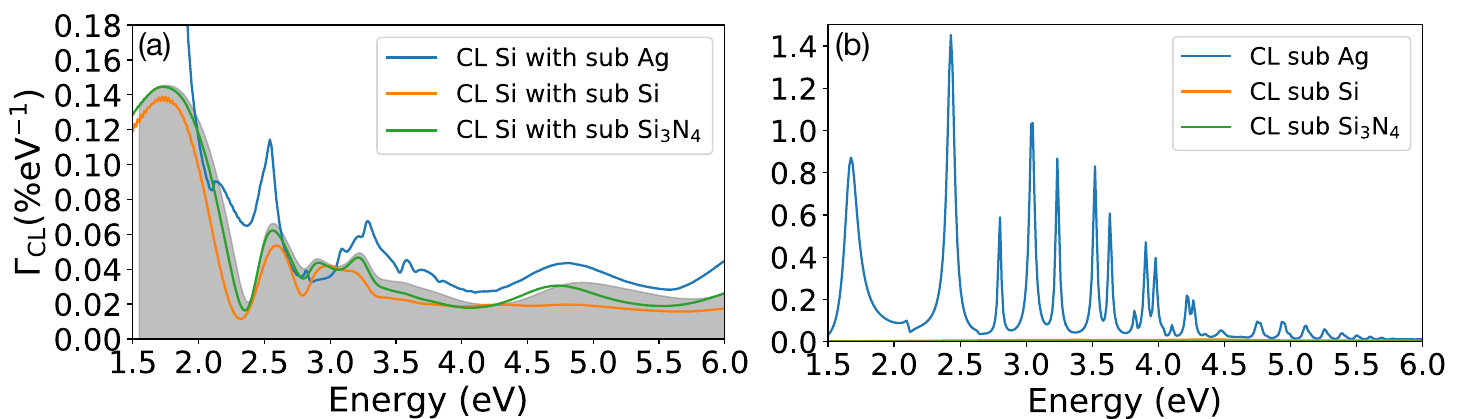}
    \caption{\textbf{CL spectra for a Si sphere deposited on a substrate of different materials.} 
    \textbf{\pnl{a}} The CL spectra for a Si sphere of radius $R=79$\,nm with different substrate materials: Si, which is implemented as numerical fitting via 5 Lorentz poles according to Eq.~\eqref{eq:Drude_lorentz}, $\mathrm{Si_3N_4}$ with constant permittivity $\varepsilon=4$, and Ag modeled as a local Drude permittivity $\varepsilon(\omega) = 1-\omega_\mathrm{p}^2/\omega(\omega+\mathrm{i}\gamma)$ with bulk plasmon energy $\hbar\omega_\mathrm{p} = 9.17$\,eV, damping constant $\hbar\gamma = 21$\,meV.
    \textbf{\pnl{b}} The CL spectra only for substrates of different materials without presence of a Si sphere.
    }
    \label{fig:Si_diff_subs}
\end{figure}

During the DG computations, we record the CL spectra on the two TF/SF contours, and the sum of them gives the total CL signal. In Fig.~\ref{fig:Si_diff_subs}\pnl{a}, we display the CL spectra for different substrate materials, where the gray color denotes the CL spectrum for a free-standing Si sphere. We note that the influence of $\mathrm{Si_3N_4}$ is relatively small compared with other substrate materials, making it a popular substrate in typical EELS/CL experiments. Remarkably, the presence of a Si substrate smears the two peaks at 2.9\,eV and 3.2\,eV, making them difficult to distinguish. Moreover, the total CL in this case is lower than in the case without a substrate. In contrast to dielectric substrates, the Ag substrate changes the CL spectrum dramatically, affecting both the peak positions and the number of peaks.
We clearly observe the splitting/emergence of peaks between 3.0\,eV and 4.0\,eV, which results from strong SPP modes supported by the Ag substrate, as shown in Fig.~\ref{fig:Si_diff_subs}\pnl{b}. There, we present the EEL spectra for the interaction between a swift electron and the substrates without the nanoparticle. We observe strong electron-induced excitations for the Ag substrate, whereas dielectric substrates provide much weaker field confinement and thus significantly reduced guided-mode excitations.

Therefore, we conclude that metal substrates are generally unsuitable for probing the intrinsic excitation properties of nanoparticles. Their high optical density of states, strong damping, and efficient coupling to SPP modes strongly perturb the local electromagnetic environment, redirecting the excitation phenomena induced by the swift electron into resonances associated with the substrate rather than the nanoparticle under investigation.

\section{Video via DGTD}

The file "Movie$\_$DGTD.mp4" shows the temporal evolution of the amplitude of the electric field $|\mathbf{E}^\mathrm{ind}(\mathbf{r},t)|=[(E^\mathrm{ind}_x)^2+(E^\mathrm{ind}_y)^2+(E^\mathrm{ind}_z)^2]^{1/2}$ in the $xz$-plane, plotted in log-scale. The swift electron beam with speed $\beta=0.7$ penetrates through a dielectric sphere of $R=79$\,nm modeled by a constant permittivity with $\varepsilon=16$. The position of the electron is indicated as a red point following the $z$-axis with impact parameter $b=0$. 

\hfill

\bibliographystyle{achemso}
\bibliography{refs}